\documentclass[aps,prd,twocolumn,floatfix,nofootinbib,superscriptaddress,tightenlines]{revtex4}

\usepackage[dvipsnames]{xcolor}
\usepackage{ragged2e}
\usepackage{soul}
\usepackage{amsmath}
\usepackage{caption}
\usepackage{dcolumn}
\usepackage{lipsum}
\usepackage{amssymb}
\usepackage{url}
\usepackage{epsfig}
\usepackage{graphicx}
\usepackage{bm}
\usepackage{setspace}
\usepackage{lscape}
\usepackage{amsthm}
\usepackage{bbold}
\usepackage{dcolumn}
\usepackage{epsfig}
\usepackage{graphics}
\usepackage{graphicx}
\usepackage{xcolor}
\usepackage{subcaption}
\usepackage[justification=raggedright,singlelinecheck=false]{caption}

\usepackage{etoolbox}
\makeatletter
\patchcmd{\@makecaption}
  {\@tempdima\hfil}
  {\@tempdima\z@}
  {}{}
\makeatother
\usepackage{bm}
\usepackage{xspace}
\usepackage{cancel}
\usepackage{float}
\usepackage{multirow}
\definecolor{darkgreen}{rgb}{0,0.5,0}
\definecolor{purple}{rgb}{0.5,0,0.5}
\definecolor{nblue}{rgb}{0.0,0.0,0.50}
\definecolor{scarlet}{rgb}{1.0,0.2,0}
\definecolor{darkmagenta}{rgb}{0.55, 0.0, 0.55}
\definecolor{darkolivegreen}{rgb}{0.33, 0.42, 0.18}
\definecolor{darkcandyapplered}{rgb}{0.64, 0.0, 0.0}

\usepackage[colorlinks=true, pdfstartview=FitV, linkcolor=purple, citecolor= purple, urlcolor=blue]{hyperref}
\usepackage[normalem]{ulem}

\newcommand{\be}{\begin{equation}}
\newcommand{\tu}{\textcolor{black}{u}}

\newcommand{\g}{\textcolor{black}{\Gamma}}
\newcommand{\D}{\textcolor{black}{\Delta}}
\newcommand{\td}{\textcolor{black}{d}}
\newcommand{\tb}{\textcolor{black}{b}}
\newcommand{\tc}{\textcolor{black}{c}}
\newcommand{\ts}{\textcolor{black}{s}}
\newcommand{\ee}{\end{equation}}
\newcommand{\bea}{\begin{eqnarray}}
\newcommand{\eea}{\end{eqnarray}}
\newcommand{\beas}{\begin{eqnarray*}}
\newcommand{\eeas}{\end{eqnarray*}}
\newcommand{\nn}{\nonumber}

\newcommand{\tq}{\textcolor{black}{q}}
\newcommand{\tqu}{\textcolor{black}{q_1}}

\newcommand{\MeV}{\text{MeV}} 
\newcommand{\GeV}{\text{GeV}} 

\newcommand{\eqn}[1]{Eq.~(\ref{#1})}

\begin{document}
\title{Exploring Screening Masses of Light and Heavy Spin-$3/2$ Baryons in a Symmetry Preserving Contact Interaction Model}

\author{R. J. Hern\'andez-Pinto}
\email{roger@uas.edu.mx}
\affiliation{Facultad de Ciencias Naturales y Exactas, Universidad Aut\'onoma de Sinaloa, Ciudad Universitaria, Culiac\'an, Sinaloa 80000,
M\'exico}

\author{J. A. Alvarado-Morales}
\email{josealvarado.fcfm@ms.uas.edu.mx}
\affiliation{Facultad de Ciencias Naturales y Exactas, Universidad Aut\'onoma de Sinaloa, Ciudad Universitaria, Culiac\'an, Sinaloa 80000,
M\'exico}

\author{M. C. Díaz-Lucas}
\email{mariadiaz.fcfm@ms.uas.edu.mx}
\affiliation{Facultad de Ciencias Naturales y Exactas, Universidad Aut\'onoma de Sinaloa, Ciudad Universitaria, Culiac\'an, Sinaloa 80000,
M\'exico}

\author{L. X. Guti\'errez-Guerrero}
\email{lxgutierrez@secihti.mx}
\affiliation{SECIHTI-Mesoamerican Centre for Theoretical Physics,
Universidad Aut\'onoma de Chiapas, Carretera Zapata Kilómetro~4, Real
del Bosque (Ter\'an), Tuxtla Guti\'errez, Chiapas 29040, M\'exico}




\begin{abstract}
We present a study of the screening masses of spin-$3/2$ baryons within a 
symmetry preserving contact interaction framework. Extending previous 
finite-temperature investigations of light baryons, we analyze 16 systems 
composed of $u$, $s$, $c$, and $b$ quarks, including light, heavy-light, 
and fully-heavy configurations with both positive and negative parity 
states. Whenever available, our results are compared with previous contact 
interaction calculations and other theoretical predictions. Our analysis 
shows that the screening masses generally increase with temperature, while 
the thermal evolution of the relative splitting between positive and 
negative parity states exhibits a clear dependence on the flavor content. 
Although a reduction of the relative parity splitting is observed in most 
channels, this behavior is not universal. Nevertheless, the uncertainties 
associated with the screening masses allow the parity partners to approach 
more closely at high temperatures, remaining compatible with an enhanced 
tendency toward parity degeneracy. Thermal modifications are strongest in 
the light sector and become increasingly suppressed as the heavy-quark 
content grows. To characterize this evolution, we introduce characteristic 
temperatures associated with the onset of thermal effects and find a general 
tendency for strong thermal modifications to occur at comparable or slightly 
higher temperatures in the negative parity channel. We further analyze the 
corresponding Faddeev amplitudes and show that the dominant diquark 
correlations remain clearly identifiable throughout the temperature range 
considered. These results provide a unified description of light and heavy 
spin-$3/2$ baryons and offer new insight into the flavor dependence of their 
thermal evolution and the persistence of quark--diquark correlations in hot 
QCD matter.
\end{abstract}


\maketitle

\section{Introduction}
Hadron physics under extreme conditions remains a vibrant area of research owing to its central role in understanding strongly interacting matter and the formation of the quark-gluon plasma (QGP). Within this framework, the investigation of quark bound states plays a crucial role in revealing how hadronic properties evolve with temperature. In particular, recent studies have highlighted the importance of exploring the behavior of baryons containing strange and heavy quarks in hot and dense media, as such systems provide valuable insights into nonperturbative aspects of quantum chromodynamics (QCD) and the structure of strongly interacting matter under extreme conditions~\cite{Messchendorp:2025men}.\\
While finite-temperature studies of mesons are abundant in the literature~\cite{Cheng:2010fe,Mukherjee:2008tr,Bazavov:2020teh,Petreczky:2009at,Song:1993ipa,Chen:2024emt,Wang:2013wk,Ramirez-Garrido:2025rsu,Gutierrez-Guerrero:2026hhu,Sumit:2023hjj,Bazavov:2019www,Hashimoto:1992np}, considerably less attention has been devoted to baryons, especially those containing heavy quarks. Heavy flavors provide unique probes of the QGP since they are predominantly produced during the earliest stages of heavy-ion collisions and remain present throughout the evolution of the medium~\cite{Rapp:2018qla,Dong:2019byy}. Consequently, baryons containing charm and bottom quarks, including fully-heavy spin-$3/2$ systems such as the $\Omega^{++}_{\tc\tc\tc}$, have attracted increasing interest as potential probes of hot QCD matter and heavy-quark dynamics~\cite{Adamova:2019vkf,Li:2025nca}. Understanding their behavior at $T\neq0$ MeV is therefore of considerable importance. Nevertheless, even at very high temperatures, the perturbative description of QCD remains limited by the infrared problem~\cite{Linde:1980ts}, highlighting the need for nonperturbative approaches. In this regard, lattice QCD calculations have shown that, at vanishing baryon chemical potential, hadronic matter undergoes a rapid crossover to a deconfined QGP accompanied by the restoration of chiral symmetry~\cite{Bazavov:2011nk}.\\
At finite-temperature, hadronic excitations can be characterized through two complementary quantities: pole masses and screening masses. Pole masses are associated with the temporal structure of correlation functions and provide information about the energy required to excite a hadronic state in the medium. Screening masses, on the other hand, are extracted from spatial correlation functions and characterize the range over which hadronic correlations propagate through the thermal environment. While both quantities contain valuable information on the in-medium behavior of hadrons, they probe different physical aspects of the system and need not exhibit the same thermal evolution.\\
Screening masses, defined through the inverse of spatial correlation lengths, provide a natural framework for characterizing the response of strongly interacting matter at finite-temperature. They constitute sensitive probes of chiral symmetry restoration and nonperturbative QCD dynamics in the hot medium~\cite{Giusti:2024ohu}. Most investigations of screening masses have focused on mesonic channels, particularly within lattice QCD~\cite{Detar:1987hib,Gocksch:1987nt}. In contrast, baryonic screening masses remain comparatively unexplored owing to the severe signal to noise problem affecting baryonic correlators, their richer spinorial structure, and the higher computational cost involved in their extraction. Moreover, their interpretation is more subtle since screening masses characterize spatial correlation lengths rather than pole masses and approach the free theory limit, $3\pi T$, only at sufficiently high temperatures~\cite{Eletsky:1988an}.\\
Thermal modifications of decuplet baryons have been analyzed previously using QCD sum rules~\cite{Azizi:2016ddw,Xu:2015jxa}. In addition, lattice-QCD calculations of octet and decuplet baryons have revealed a progressive reduction of the mass splitting between positive and negative parity states and the emergence of parity doubling near the crossover region~\cite{Aarts:2018glk}. However, these works focused on temporal masses rather than screening masses extracted from spatial correlation functions. Consequently, the thermal evolution of spin-$3/2$ baryonic screening masses remains largely unexplored.
Accordingly, nonperturbative approaches such as the symmetry preserving contact interaction (CI) model~\cite{GutierrezGuerrero:2010md} provide a valuable framework for investigating hadronic properties beyond the reach of perturbative methods. Over the years, the CI approach has successfully described a wide variety of hadronic observables, including ground and excited states of mesons and baryons, as well as the finite-temperature behavior of scalar, pseudoscalar, vector (V), and axial-vector (AV) mesons and their corresponding diquark correlations. Previous finite-temperature studies within this framework have focused predominantly on light-baryon systems. Here, we extend these investigations to spin-$3/2$ baryons spanning the light, strange, charm, and bottom sectors. To the best of our knowledge, this constitutes the first study of the screening masses of heavy and fully-heavy spin-$3/2$ baryons at finite-temperature. These systems offer a unique opportunity to examine how chiral symmetry restoration, confinement-related effects, and heavy-quark dynamics manifest themselves in strongly interacting matter under extreme thermal conditions.
This paper is organized as follows. In Sec.~\ref{CI}, we briefly review the CI framework and summarize the ingredients required for the present study. The formulation of the finite-temperature Faddeev equation (FE) for spin-$3/2$ baryons is presented in Sec.~\ref{Baryons}, where positive and negative parity states are discussed separately in Subsections~\ref{secc-p} and~\ref{secc-n}, respectively. The resulting screening masses and Faddeev amplitudes (FAs) are analyzed in Sec.~\ref{Results}. Finally, our conclusions are presented in Sec.~\ref{Summary}. Additional details concerning the notation employed throughout this work and the flavor matrices associated with the different diquark configurations are collected in Appendices~\ref{App:EM} and~\ref{App:Flavdiq}, respectively.
\section{Background on the CI Model} \label{CI}
In this section, we briefly review CI framework and introduce the key elements relevant to the present study. In particular, we discuss the finite-temperature gap equation, the model parametrization, the temperature dependence of the dressed-quark masses, and the notation adopted throughout this work. These ingredients constitute the foundation for the calculation of the baryon screening masses presented in the following sections.
\subsection{The Gap Equation at Finite-Temperature}
\label{A}
\begin{figure}[b]
   \vspace{-2cm}
   \centering
    \includegraphics[scale=0.3,angle=-90]{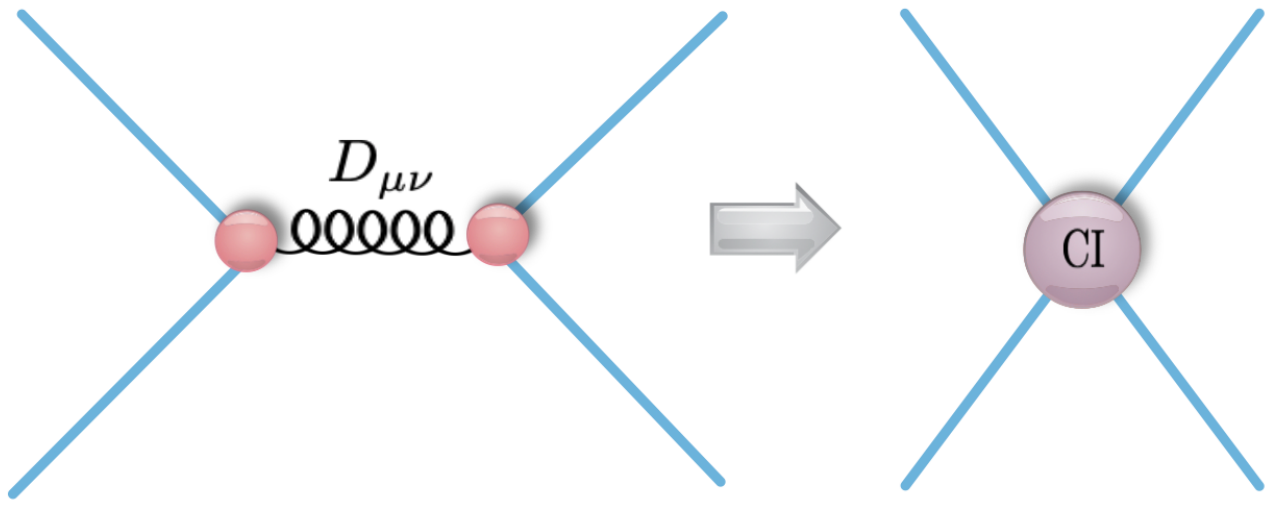}
    \vspace{-2cm}
    \caption{\justifying Illustration of the CI framework. The left panel shows the full momentum-dependent gluon propagator, whereas the right panel displays the momentum-independent approximation defined in Eq.~(\ref{eqn:contact_interaction}).}
    \label{fig:ci}
\end{figure}

In analogy with the temperature-independent case, the core of our study for $T\neq 0$ MeV lies in the dressed-quark propagator for a quark of flavor $f$, which is determined by solving the gap equation,
\begin{align}
    \nn S_f^{-1}(p;T) &=i \vec{\gamma} \cdot \vec{p} +i \gamma_4 \omega_n +m_f \\
    &+ \frac{16 \pi \hat{\alpha}_{\mathrm{IR}}}{3} \int_{\ell,dq} \gamma_{\mu} S_f(q;T)\gamma_{\mu}\,,
\end{align}
where $p^{\mu}=(\omega_n,\vec{p})$ and $q^{\mu}=(\omega_\ell,\vec{q})$ are the four momenta of the quarks, $f$ represents the quark flavor, $m_f$ the current mass of the quark, and $\omega_n =(2n+1)\,\pi \,T$ is the fermion Matsubara frequency. Besides, we have used the notation,
\bea\int_{\ell,dq}:=T\sum^{\infty}_{\ell=-\infty}\int \frac{d^3\vec{q}}{(2\pi)^3}\,.\eea
For the gluon propagator, we have implemented the  CI approach, i.e., 
\begin{eqnarray}
\label{eqn:contact_interaction}
g^{2}D_{\mu \nu}(k)&\to&4\pi\,\hat{\alpha}_{\mathrm{IR}}\,\delta_{\mu \nu}  \,,
\end{eqnarray}
\noindent  where $\hat{\alpha}_{\mathrm{IR}}=\alpha_{\mathrm{IR}}/m_g^2$, with $\alpha_{\mathrm{IR}}$ the running-coupling and \,$m_g=500\,\MeV$  the gluon mass scale~\cite{Aguilar:2017dco,Binosi:2017rwj,Gao:2017uox}.   A diagrammatic representation of the model is shown in Fig.~\ref{fig:ci}.\\
The finite-temperature dressed-quark mass $M_f(T)$ is then extracted by means of
\begin{align}
\label{dressed mass}
    M_f(T) =m_f +\frac{16 \pi \hat{\alpha}_{\mathrm{IR}}}{3} \int_{\ell,dq} \frac{4M_f(T)}{s_{\ell}+M_f(T)^2}  \, ,
\end{align}
    with $s_\ell=\vert\vec{q}\vert^2+\omega^2_\ell$. Using the Poincaré invariant method, we find,
    \begin{align}\label{gap-T}
        M_f(T) =m_f +M_f(T)\frac{4  \hat{\alpha}_{\mathrm{IR}}}{3\pi} \mathcal{C}^{\rm iu} (M_f(T)^2;T) 
    \end{align}
    where
    \begin{align}
        \nn \mathcal{C}^{\rm iu} (M_f(T)^2;T) = 2 \sqrt{\pi} \, T \, \int_{\tau_{\rm UV}^2}^{\tau_{\rm IR}^2} d\tau \, \frac{e^{- M_f^2 \, \tau} \vartheta_2\left(e^{- 4\pi^2 T^2\, \tau}\right)}{\tau^{3/2}}\,,
    \end{align}
with $\vartheta_2(x)$ the Jacobi theta function
and 
${\cal C}^{\rm iu}(\sigma;T)/\sigma = \overline{\cal C}^{\rm iu}(\sigma;T)\,$.
$\tau_{\rm IR, UV}$ are respectively, infrared and ultraviolet regulators or, similarly, $\Lambda_{\rm IR,UV} = \tau_{\rm IR, UV}^{-1}$.  
\subsection{Parameters}
\label{B}
For our calculations, we adopt the parameters introduced in Ref.~\cite{Gutierrez-Guerrero:2021rsx}, where a unified framework was employed to compute the masses of mesons and baryons composed of both light and heavy quarks in CI model. These parameters were also used to describe successfully the parity partners and are listed in Table~ \ref{parameters}.
 \begin{table}[b]
 \caption{\justifying
\label{parameters}
Parameters defining the interaction for the different quark sectors.
The effective interaction strength is given by
$\hat{\alpha}_{\mathrm{IR}}=\hat{\alpha}_{\mathrm{IRL}}/Z_H$, where
$\hat{\alpha}_{\mathrm{IRL}}=4.57$.
Here, $Z_H$ denotes the heavy-quark suppression factor introduced in Ref.~\cite{Gutierrez-Guerrero:2021rsx}, and
$\Lambda_{\mathrm{UV}}$ is the ultraviolet regularization scale.
The infrared regulator is fixed at
$\Lambda_{\mathrm{IR}}^{(0)}=0.24~\mathrm{GeV}$ for $T=0$ MeV and evolves according to Eq.~(\ref{irt}) at finite-temperature.}
\begin{center}
\label{parameters1}
\begin{tabular}{@{\extracolsep{0.0 cm}} || l | c | c | c ||}
\hline \hline
 \, quarks \, &\,  $Z_{H}$ \, &\,  $\Lambda_{\mathrm {UV}}\,[\GeV] $ \,  &\,  $\hat{\alpha}_{\mathrm{IR}}\,[\GeV^{-2}]$
 \\
 \hline
 \rule{0ex}{2.5ex}
$\, \tu,\td, \ts$ & 1 & 0.905 & 4.57   \\ 
\rule{0ex}{2.5ex}
$\, \tc, \td, \ts $ & \, 3.034 \, & 1.322 & 1.51 \\ 
\rule{0ex}{2.5ex}
$\, \tc $ & \, 13.122 \, & 2.305 & 0.35 \\ 
\rule{0ex}{2.5ex}
$\,  \tb,\tu$, \ts & \, 16.473 \, & 2.522 & 0.28 \\ 
\rule{0ex}{2.5ex}
$\, \tb, \tc$     &  59.056 & 4.131 & 0.08 \\
\rule{0ex}{2.5ex}
$\, \tb $ & 165.848 & 6.559 & 0.03\\
\hline \hline
\end{tabular}
\end{center}
\end{table}
The values in Table \ref{parameters} indicate that $\hat{\alpha}_{\mathrm {IR}}$ naturally varies with the mass scale $\Lambda_{\mathrm{UV}}$ as described in Ref.~\cite{Bedolla:2015mpa} and subsequently employed in various later studies~\cite{Bedolla:2016yxq, Raya:2017ggu, Gutierrez-Guerrero:2019uwa, Yin:2019bxe, Yin:2021uom}. 
It is important to highlight that, in the calculation of hadron masses in their ground and excited states  at $T=0$ MeV, we used a constant $\Lambda^{(0)}_{\rm IR} = 0.24$ GeV, since this infrared regulator implements confinement by
ensuring the absence of quark production thresholds in
all processes. However, for screening masses, this value must vary with temperature \cite{Chen:2024emt,Wang:2013wk}, that is,
$\Lambda^{(0)}_{\rm IR} \to\Lambda_{\rm IR}(T)\,.$
We followed previous works \cite{Chen:2024emt,Wang:2013wk} and defined  the infrared regulator evolution as,
\bea
\label{irt}
\Lambda_{\rm IR}(T) =\Lambda_{\rm IR}^{(0)} \left(\frac{ M_f(T)}{M_f(0)}\right)^{1/4} \, ,
\eea
 where the value of $\Lambda_{\rm IR}^{(0)}$
  corresponds to the constant infrared scale used at $T=0$ MeV, namely $0.24$ GeV.
With all these parameters in hand, the following subsection explores the behavior of the dressed-quark masses for the $\tu$, $\ts$, $\tc$, and $\tb$ flavors\footnote{We include $\td$-quarks by imposing isospin symmetry.} at nonzero temperatures. Where applicable, our results are compared with those obtained from more sophisticated approaches.
\subsection{Finite-Temperature Dressed-Quark Masses}
Table \ref{table-M} establishes the current quark masses used in our calculations at $T=0$ MeV; however, we now compute the dressed-quark masses as a function of the temperature using the parameters presented above.
Table~\ref{table-M} presents the results of our calculation using \eqn{dressed mass} for the dressed-quark mass at zero temperature.

\begin{table}[h!]
\caption{\justifying \label{table-M}
Current ($m_{f}$) and dressed-quark masses
($M_{f}$) for quarks in GeV obtained using $T=0$ MeV.}
\begin{center}
\begin{tabular}{@{\extracolsep{0.0 cm}} || c | c | c | c || }
\hline 
\hline
 $m_{\tu}=0.007$ &$m_{\ts}=0.17$ & $m_{\tc}=1.08$ & $m_{\tb}=3.92$   \\
 \rule{0ex}{2.5ex}
 $ M_{\tu}=0.367$ \, & \, $  M_{\ts}=0.53$\; \, &\,   $  M_{\tc}=1.52$ \, &\,  $  M_{\tb}=4.75$   \\
 \hline
 \hline
\end{tabular}
\end{center}
\end{table}
\begin{figure}[b]
    \centering    \includegraphics[width=1\linewidth]{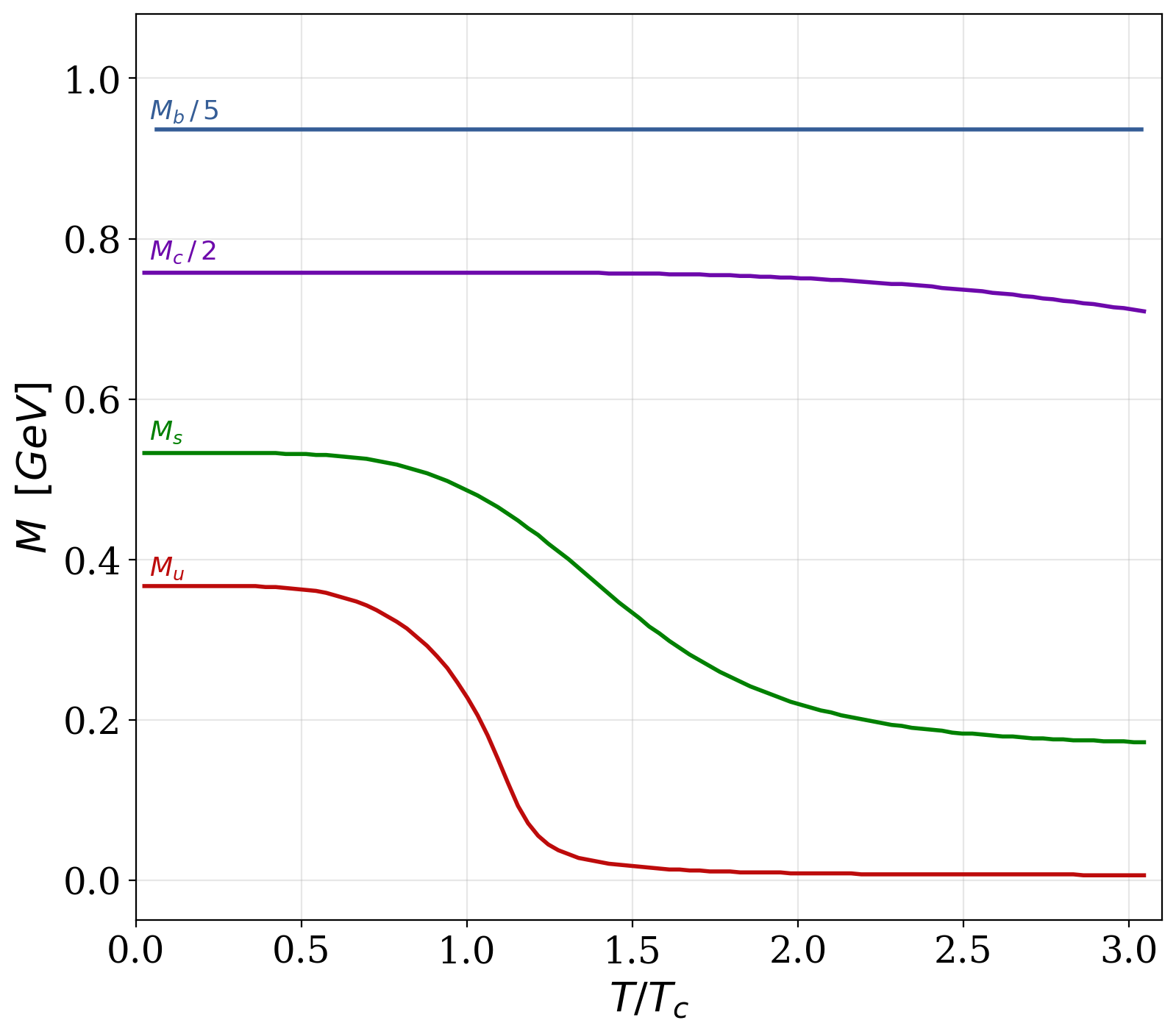}
    \vspace{-0.7cm}
  \caption{\justifying
Temperature dependence of the finite-temperature dressed-quark masses
within the CI framework for $T_c=197~\mathrm{MeV}$~\cite{Ramirez-Garrido:2025rsu,Gutierrez-Guerrero:2026hhu}.
The results are presented as a function of the dimensionless
temperature ratio $T/T_c$ and extend up to $T/T_c=3$.
At sufficiently high temperatures, the dressed-quark masses
asymptotically approach their corresponding current masses,
$M_{f}(T)\rightarrow m_{f}$.}
    \label{fig:dressed}
\end{figure}
We illustrate the evolution of the dressed-quark masses as a function of temperature in Fig.~\ref{fig:dressed}.
In the present work, which includes light, heavy-light, and fully-heavy baryons, we adopt $T_c = 197$ MeV, consistent with previous studies within the CI framework. This value is larger than the pseudocritical temperature typically employed in lattice QCD simulations, $T_c \approx 155$ MeV \cite{Andronic:2017pug, Bazavov:2011nk}. The use of $T_c = 197$ MeV ensures consistency with earlier CI calculations and facilitates comparison with related studies. Nevertheless, within the present framework $T_c$ functions primarily as a reference scale, such that its numerical value acts essentially as a rescaling factor rather than influencing the qualitative thermal behavior of the system.
It is worth highlighting that, as the temperature rises, the dressed-quark mass decreases, a trend that is particularly pronounced for light quarks.
We make this observation more explicit in Table~\ref{T600}, where the current andfinite-temperature dressed-quark masses are compared at $T=500$ MeV. The largest relative deviation is observed for the charm quark,
reaching 25.4\% with respect to the dressed-quark mass,
followed by the bottom quark with a deviation of 16.2\%.
For lighter quarks, the discrepancies remain
below 3.5\%.
\begin{table}[t]
    \centering
  \caption{\justifying
Finite-temperature dressed-quark masses
at $T=500$ MeV in comparison with current masses.
Here, $m_f^{\mathrm{CI}}$ denotes the current quark mass
employed in the CI model. All masses are in GeV.}
    \label{T600}
    \begin{tabular}{@{\extracolsep{0.0 cm}}c|c|c}
    \hline\hline
    \, Quark \, & \, $m_f^{\rm CI}$ \, & \, $M_{f}(T=500\text{ MeV})$ \, \\
    \hline
       $\tu$, $\td$   & 0.007 & 0.007 \\
       $\ts$   &  0.17& 0.176\\
       $\tc$    &  1.08& 1.447\\
       $\tb$  &  3.92 & 4.676\\ 
       \hline\hline 
    \end{tabular}
\end{table}
In addition, we point out that the behavior of the dressed-quark masses as a function of
temperature is as expected. For instance, for the $\tu$ and $\ts$ quarks, we compare our results with those obtained using the NJL model \cite{Sheng:2022ssp} in Fig. \ref{fig:compa-njl} showing the correct trend.\\
\begin{figure}[b]
    \vspace{-0.cm}
    \centering
    \hspace{-0.6cm}
    \includegraphics[width=1.05\linewidth]{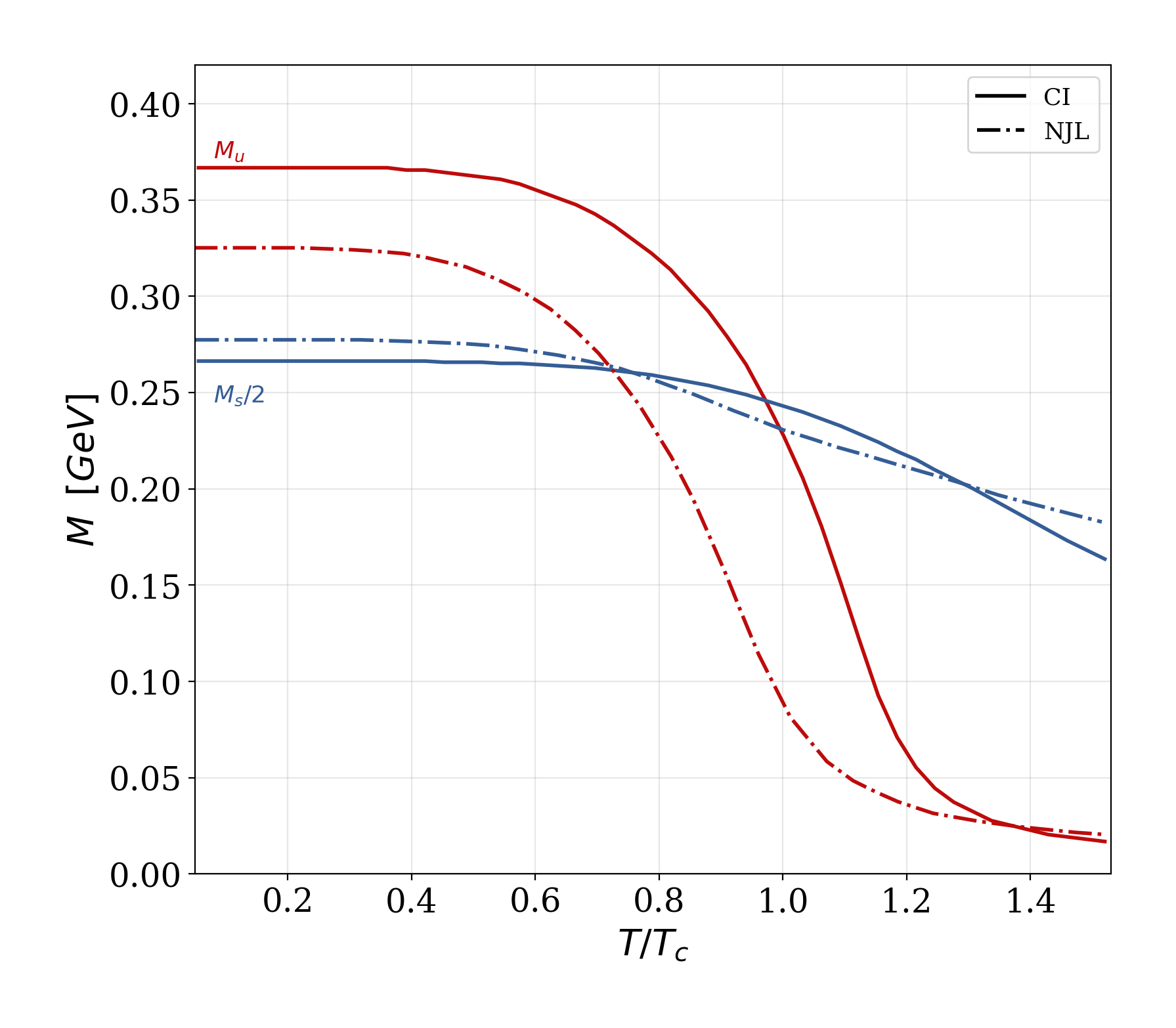}
    \vspace{-1cm}
\caption{\justifying Finite-temperature dressed-quark masses obtained in this work (solid lines) and those derived from the NJL model (dash-dotted lines) from Ref.~\cite{Sheng:2022ssp}. In this figure, we use $T_c=197$ MeV.}

    \label{fig:compa-njl}
\end{figure}
Once the CI framework and the thermal behavior of quarks have been established, we proceed to investigate three-quark bound states, with particular emphasis on the spin-$3/2$ sector. In the next section, we introduce the corresponding Faddeev equation (FE) within the CI model and discuss the ingredients required to compute the associated screening masses, closely following the notation of Ref.~\cite{Gutierrez-Guerrero:2019uwa}.
\section{Faddeev Equation}
\label{Baryons}
\begin{figure}[b]
\begin{center}
\label{figura3}
\includegraphics[width=7.5cm]{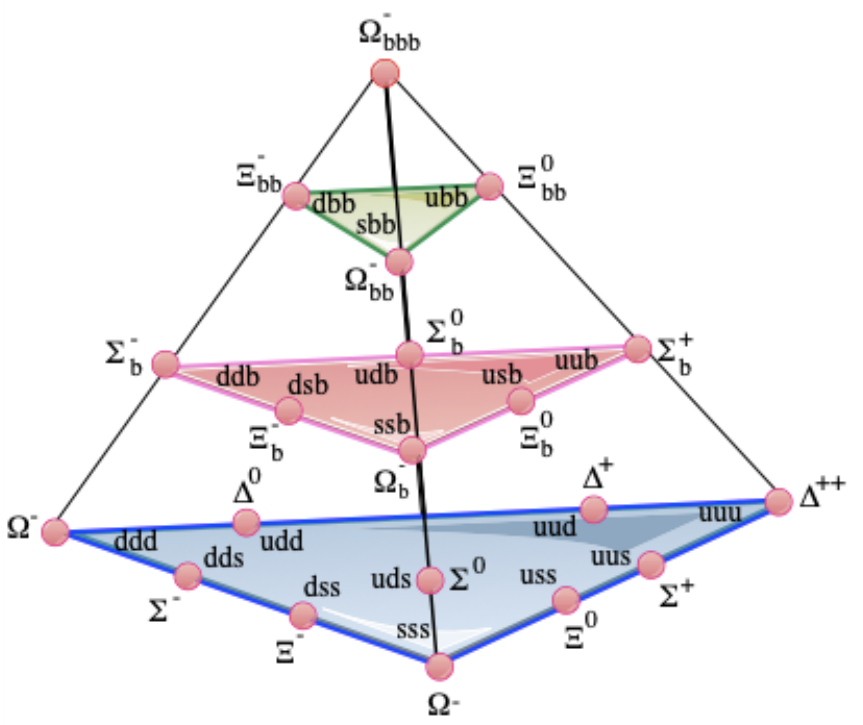}
\caption{\label{octydec} \justifying Baryons with spin-$3/2$ made from four quarks of the types $\tu, \td, \ts$, and $\tb$. Doubly heavy baryons and triply heavy baryons are localized in the highest layers.}\label{figures}
\end{center}
\end{figure}

\begin{figure}[htpb]
\vspace{-3.0cm}
\hspace{-5mm}    \centerline{\includegraphics[scale=0.40,angle=-90]{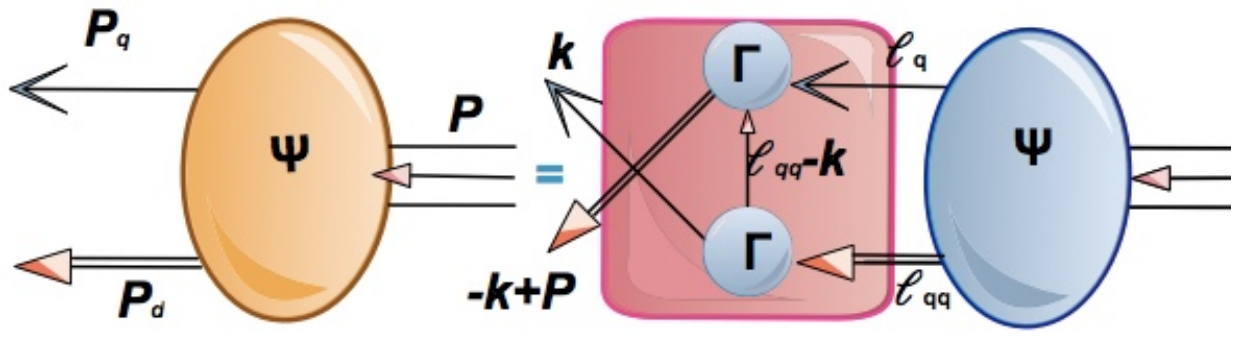}}       \vspace{-2.8cm}\caption{\justifying Poincar\'e-covariant Faddeev equation for the baryon screening masses studied in this work. The red square represents the quark-diquark interaction Kernel. The single line denotes the dressed quark, the double line is the diquark while $\Gamma$ and $\Psi$ are the BSA and FA, respectively. Configuration of momenta is: $\ell_{qq}=-\ell + P$, $k_{qq}=-k+P$, $P=P_d+P_q$.}
       \label{faddevv-Fig1}
\end{figure}

We now focus on the baryon sector. Since the systems considered here span the light, strange, charm, and bottom sectors, it is convenient to organize them according to their flavor structure. The flavor decompositions $3\otimes3\otimes3$ and $4\otimes4\otimes4$ generate the familiar $SU(3)$ and $SU(4)$ baryon multiplets, respectively. Although explicit quark masses break the underlying flavor symmetries, this classification remains useful for organizing the states studied in the present work. Figure~\ref{octydec} shows representative baryon multiplets; the charm sector is analogous to the bottom sector.\\
%
In the present work, baryons, viewed as bound states of three quarks, are described within the quark-diquark picture, where the original three-body problem is reduced to an effective two-body system. Within this framework, the dynamics are governed by the FE, illustrated in Fig.~\ref{faddevv-Fig1}. This approach has proven to be a reliable and computationally efficient framework for investigating baryon properties in continuum studies of QCD \cite{Wilson:2011aa}.
Spin-$3/2$ baryons provide a valuable laboratory for investigating the interplay between flavor, spin, and confinement in QCD. In particular, they include systems containing up to three charm or three bottom quarks, offering direct access to heavy-quark dynamics in a strongly interacting environment. Their internal structure is also subject to stringent constraints: since a scalar diquark combined with a spin-$1/2$ quark cannot produce a state with total spin-$3/2$, these baryons are necessarily governed by spin-1 diquark correlations, namely AV and V diquarks.
The present study builds on Refs.~\cite{Ramirez-Garrido:2025rsu,Gutierrez-Guerrero:2026hhu}, where the temperature dependence of diquark screening masses and Bethe Salpeter Amplitudes (BSAs) was determined within a CI framework. Using these quantities as input to the finite-temperature FE, we investigate the screening masses of decuplet baryons and extend the CI approach to states containing $\tc$ and $\tb$ quarks, thereby providing a unified description of light- and heavy-flavor sectors.
\\
A key advantage of the CI framework is the substantial reduction in complexity. While a fully momentum-dependent three-body description requires $64$ Dirac structures for the Faddeev Amplitudes (FAs)~\cite{Eichmann:2009en}, only $8$ independent tensors remain in the quark-diquark approximation. For spin-$3/2$ baryons, the structure simplifies further because only  V and AV diquark correlations contribute. The resulting bound state emerges through repeated quark exchange between the diquark and spectator-quark sectors. The validity of this approximation has been established in Ref.~\cite{Eichmann:2009qa}, where it reproduces the nucleon mass at  approximately 5\% level of accuracy. 
The decuplet baryons considered in this work, including both positive and negative parity states, are described within a unified Poincar\'e-covariant quark-diquark framework. Accordingly, their general FA at finite-temperature can be expressed as: 
\bea
\Psi_\mu^{\pm} = \mathcal{D}_{\mu\nu}^{\pm}(P;T)\,u_\nu(P,s) \;, \nn 
\eea
where $P$ is the total momentum of the baryon, $s$ labels the spin
projection, and $u_\nu(P,s)$ denotes the corresponding Rarita--Schwinger spinor. The tensor structure of $D_{\mu\nu}^{\pm}(P;T)$ is determined by the spin-$3/2$ nature of the baryon and can be written as
\bea
&& \mathcal{D}_{\mu\nu}^{\pm}(P;T)
= f^{\pm}(P;T)\,\delta_{\mu\nu}\,G^{\pm},
\label{base}
\eea
and $G^{+(-)} = \mathbb{I}_{\rm D}(\gamma_{5})$. The explicit structure of ${\cal D}_{\mu\nu}^{\pm}(P;T)$ is determined by the quark-flavor content of the baryon, as detailed in the following sections.\\
In the CI model, the bound-state BSAs for V and AV diquarks composed of quarks $\tq$ and $\tqu$ have the form:
\begin{align}
\nn \Gamma_{\{\boldsymbol{\cdot}\},\mu}^{1^-} (P;T)
&=\gamma_5
\gamma_{\mu}^{\perp}\, E_{\{\boldsymbol{\cdot}\}}^{1^-},
\qquad
\gamma_{\mu}^{\perp} P_{\mu} = 0,
\\[6pt]
\nn \Gamma_{\{\boldsymbol{\cdot}\},\mu}^{1^+}(P;T)
&=
\gamma_{\mu}^{\perp}\, E_{\{\boldsymbol{\cdot}\}}^{1^+} +
\frac{1}{2\,M_R(T)}
\,\sigma_{\mu\nu} P_{\nu}\,
F_{\{\boldsymbol{\cdot}\}}^{1^+}.
\end{align}
with $M_R(T)=M_{\tqu}M_{\tq}/(M_{\tqu}+M_{\tq})$ the corresponding reduced mass. Here, we introduce the notation ${\boldsymbol{\{\cdot\}}}$ as a generic 
placeholder for the flavor content of the diquark, thus avoiding the explicit 
specification of the constituent quark flavors in the BSA labels. For 
notational simplicity, we also suppress the explicit momentum and temperature 
dependence of the amplitudes and write
\begin{equation}
X_{{\boldsymbol{\{\cdot\}}}}^{1^{\pm}}
\equiv
X_{{\boldsymbol{\{\cdot\}}}}^{1^{\pm}}(P;T),
\qquad X=E,F,
\end{equation}
whenever applicable. The explicit temperature dependence of the dressed-quark 
masses is likewise omitted throughout. In this notation, 
$E_{{\boldsymbol{\{\cdot\}}}}^{1^{\pm}}$ and 
$F_{{\boldsymbol{\{\cdot\}}}}^{1^+}$ denote the corresponding canonically 
normalized BSAs.
The quantities $\Delta_{\{\boldsymbol{\cdot}\},\mu\nu}^{1^\pm}(K;T)$ 
denote the standard finite-temperature propagators associated with the V and AV 
diquark correlations and are defined as
\begin{align}
\Delta_{\{\boldsymbol{\cdot}\},\mu\nu}^{1^\pm}(K;T)
&=
\frac{1}{K^2 + m_{\{\boldsymbol{\cdot}\}, 1^\pm}^2(T)}
\left(
\delta_{\mu\nu}
+
\frac{K_\mu K_\nu}{m_{\{\boldsymbol{\cdot}\},1^\pm}^2(T)}
\right),
\end{align}
where $m_{\{\boldsymbol{\cdot}\},1^-}(T)$ and 
$m_{\{\boldsymbol{\cdot}\},1^+}(T)$ denote the temperature-dependent 
V and AV diquark masses, respectively, obtained from the corresponding 
Bethe Salpeter equations (BSEs). The derivation and analysis of the Faddeev 
equations (FEs) for the positive and negative parity states are presented in 
the following subsections.
\subsection{Positive Parity States}
\label{secc-p}
We begin our analysis with the positive parity decuplet baryons composed either of three identical quarks, $\tq\tq\tq$, or of two identical quarks accompanied by a third quark of different flavor, $\tqu\tq\tq$. Within the present framework, spin-$3/2$ positive parity baryons are constructed solely from AV diquark correlations. Consequently, we first examine the corresponding Faddeev amplitudes before discussing the resulting screening masses.\\\\
\noindent
{\bf Triply heavy and light baryons ($\tq\tq\tq$):} Only one possible type of diquarks exists for a baryon composed of same three quarks $(\tq\tq\tq)$. In this case, the FA is \cite{Gutierrez-Guerrero:2019uwa}:
\begin{equation}
{\cal D}_{\mu\nu}^+(P;T) u_\nu(P,s) = f^+(P;T) \, \mathbb{I}_{\rm D} \, u_\mu(P,s)\,.
\label{DnurhoI}
\end{equation}
%
Employing Feynman rules for Fig.~\ref{faddevv-Fig1} and using the expression for the FA, Eq.~({\ref{DnurhoI}}), we can write
\begin{align}
\nn f^{+}(P;T) \, u_\mu(P,s)&=4\, S^{\rm T}(T)
\\ \nn  & \times  
\int \frac{d^4 \ell}{(2\pi)^4}\mathcal{M_{\mu\nu}}(T)f^{+}(P;T) \, u_\nu(P,s) \,,
\end{align}
where $S^{\rm T}(T)$ denotes the exchanged-quark propagator. Previous studies within the  CI framework frequently employed the so-called static approximation, in which the exchanged-quark propagator is replaced by a constant term proportional to the inverse dressed-quark mass. This renders the Faddeev kernel momentum independent and substantially reduces the numerical complexity of the problem. The reliability of this approximation is supported by the large magnitude and weak momentum dependence of the dressed-quark mass within the CI framework; see Refs.~\cite{Roberts:2011cf,Gutierrez-Guerrero:2019uwa,Gutierrez-Guerrero:2021rsx}.\\
At finite-temperature, however, the dressed-quark mass becomes medium dependent and, in the chiral limit, is expected to vanish for $T>T_c$ \cite{Ramirez-Garrido:2025rsu,Gutierrez-Guerrero:2026hhu,Chen:2024emt}. In this regime, the inverse dressed-quark mass
replacement becomes ill-defined, and the static
approximation gradually loses its validity. A natural extension consists in promoting the quantities entering the exchanged-quark propagator to temperature-dependent functions determined from the corresponding finite-temperature gap equation. Such a procedure preserves the simplicity of the original framework while incorporating the leading effects of the thermal medium.\\
Accordingly, the exchanged-quark propagator is modeled as
\bea
S^{\rm T}=\frac{g_B}{M}
\;\longrightarrow\;
S^{\rm T}(T)=\frac{g_B(T)}{M(T)}\, ,
\eea
with
\bea
g_B(T)=g_B^{(0)}
\left(
\frac{M_R(T)}{M_R(0)}
\right)^{1/2},
\label{gb-static}
\eea
where $M$ is the dressed-quark mass of the exchanged quark,
$M_R(0)\equiv M_R(T=0 \,\, \rm MeV)$ is the reduced mass
evaluated at zero temperature, $g_B$ is the effective
quark-exchange coupling appearing in the Faddeev kernel,
and $g_B^{(0)}$ denotes its value at $T=0$ MeV.

The temperature dependence introduced in
Eq.~(\ref{gb-static}) is a phenomenological prescription intended
to incorporate the thermal evolution of the effective
quark-exchange strength within the static approximation.
The reduced mass $M_R(T)$ provides a characteristic
mass scale that reflects the temperature dependence of the dressed-quark masses. The square-root dependence
allows the effective coupling to evolve moderately
with temperature while preserving its zero-temperature
value, $g_B(T=0)=g_B^{(0)}$.
The choice of the exponent $1/2$ in Eq.~(\ref{gb-static}) is further supported by a sensitivity analysis, which shows that it provides a thermal evolution of the baryon screening masses more consistent with previous results obtained within the CI model and the expected high-temperature trend associated with $3\pi T$.
For the temperature range considered in the present
work ($T/T_c\lesssim 2$), the dressed-quark masses
remain finite, although substantially reduced for
light quarks. Consequently, the finite-temperature
extension of the static approximation remains well
defined throughout the explored temperature domain.
This feature is even more pronounced in systems
containing heavy quarks, whose effective masses
exhibit a significantly weaker thermal dependence.
We now discuss the structure of the matrix $\mathcal{M}_{\mu\nu}(T)$. As previously noted, momentum arguments are suppressed, whereas the explicit temperature dependence is retained, since the thermal evolution of the system is the primary focus of this work. This convention will be adopted in the 
calculations that follow. Its form is dictated by the diquark correlations 
composing the baryon. In the present case, only a single AV diquark correlation 
contributes, and therefore the kernel reduces to a single matrix element:
\begin{align}
\nn\mathcal{M}_{\mu\nu} (T)&=2i\g^{1^+}_{\{\tq\tq\},\rho}(l_{\tq\tq};T)i\overline{\g}^{1^+}_{\{\tq\tq\},\mu}(-k_{\tq\tq};T)\\
& \times S_{\tqu}(l_{\tqu};T)\D^{1^+}_{\{\tq\tq\},\rho\nu}(l_{\tq\tq};T)\;.
\end{align}
We 
now multiply both sides by ${\bar{u}}^B_{\beta}(P,s)$ from the left and sum over the polarization, to obtain
\bea
\Lambda_{+}(P)R_{\mu\beta}(P)=4S^{\rm T}(T)\int \frac{d^4 \ell}{(2\pi)^4}\mathcal{M_{\mu\nu}}(T)\Lambda_{+}(P)R_{\nu\beta} \,.\nn \\
\label{ef}
\eea
 Positive energy projection operator $\Lambda_+$ and  $R_{\mu\beta}$ are defined in Eqs.~(\ref{Lplus}) and (\ref{Rde}) in Appendix~\ref{App:EM}.\\
To determine the screening masses, we  consider the color-singlet bound states composed of three identical quarks:
\begin{eqnarray}
&&\nn \begin{array}{cc}
u_{\Delta^{++}}= \left[
\begin{array}{c}
\{\tu\tu\} \tu \\
\end{array} \right], &
\hspace{0.85cm}u_{\Omega^-}=
\left[ \begin{array}{c}
\{\ts\ts\} \ts \\
\end{array} \right],
\end{array}
\\
&&\nn \begin{array}{cc}
u_{\Omega_{\tc\tc\tc}^{++}}= \left[
\begin{array}{c}
\{\tc\tc\} \tc \\
\end{array} \right], &
\hspace{1.0cm}u_{\Omega_{\tb\tb\tb}^{-}}=
\left[ \begin{array}{c}
\{\tb\tb\} \tb \\
\end{array} \right]~.
\end{array}
\end{eqnarray}
For simplicity, we adopt the notation of Ref.~\cite{Chen:2012qr} for the 
flavor and spin structure of the diquarks. Accordingly, $\{qq\}$ and 
$\{q_1q\}$ denote positive parity spin-1 diquarks, with the $1^+$ 
superscript omitted by convention. This notation is particularly convenient 
because, as will become clear in the following sections, only positive parity 
diquark correlations are required in the Faddeev equations for the baryon 
states considered herein. Negative parity spin-1 diquarks, whenever discussed, 
are explicitly identified by the $1^-$ superscript. This notation will be 
used throughout the remainder of the manuscript.\\
 The screening mass at temperature $T$ is obtained when the eigenvalue associated with Eq.~(\ref{ef}) reaches unity. We emphasize that the masses of both the constituent quarks and the diquark correlations evolve with temperature and must therefore be determined self-consistently  across the entire $T$ domain.  \\ \\
%
\noindent
{\bf Mixed flavor baryons ($\tqu\tq\tq$):}
 For a baryon with the quark component structure ($\tqu\tq\tq$), there are two possible AV diquarks, $\{\tq\tq\}$ and $\{\tqu\tq\}$. The FA for such a baryon is:
\begin{equation}  \label{d-diquark}
{\cal D}_{\mu\nu}^{B}(P;T)u_{\nu}^{B}(P,s) =
\sum_{i} d_{i}(P;T) \delta_{\mu\nu} u_{\nu}^{B}(P,s),
\end{equation}
where $i=\{\{\tqu\tq\},\{\tq\tq\}\}$, so that the corresponding FE has the form
\begin{equation}
\left[\begin{matrix}
d_{\{\tqu\tq\}}\\ d_{\{\tq\tq\}}\end{matrix}\right]
u_{\mu}(P,s)= \int \frac{d^{4}l}{(2\pi)^{4}} \\
\mathbb{M}_{\mu\nu}(T)
\left[\begin{matrix}
d_{\{\tqu\tq\}}\\ d_{\{\tq\tq\}}\end{matrix}\right]
u_{\nu}(P,s) \,,
\label{eq:D2}
\end{equation}%
with
\bea
\label{M32}
\mathbb{M}_{\mu\nu}(T)=-4\left[\begin{matrix}
{\cal M}_{\mu\nu}^{11}(T) & {\cal M}_{\mu\nu}^{12}(T) \\
\rule{0ex}{3.5ex}
{\cal M}_{\mu\nu}^{21}(T) & {\cal M}_{\mu\nu}^{22}(T)
\end{matrix}\right]\;,
\eea
and matrix elements given by~:
\begin{eqnarray}
\nn {\cal M}_{\mu\nu}^{11}(T)&=& t^{f_{11}} \,
\Gamma_{\{\tqu\tq\},\rho}^{1^{+}}(\ell_{\tqu\tq};T) S^{\rm T}(T) \,
\bar{\Gamma}_{\{\tqu\tq\},\mu}^{1^{+}}(-k_{\tqu\tq};T) \\
& \times& \, S(l_{\tq};T) \,
\Delta_{\{\tqu\tq\},\rho\nu}^{1^{+}}(\ell_{\tqu\tq};T), \\
\nn {\cal M}_{\mu\nu}^{12}(T)&=&t^{f_{12}} \,
\Gamma_{\{\tq\tq\},\rho}^{1^{+}}(\ell_{\tq\tq};T) S^{\rm T}(T) \bar{\Gamma}_{\{\tqu\tq\},\mu}^{1^{+}}(-k_{\tqu\tq};T) \,\\
&\times& S(l_{q_1};T) \,
\Delta_{\{\tq\tq\},\rho\nu}^{1^{+}}(\ell_{\tq\tq};T),\\
\nn {\cal M}_{\mu\nu}^{21} (T)&=&t^{f_{21}} \,
\Gamma_{\{\tqu\tq\},\rho}^{1^{+}}(\ell_{\tqu\tq};T) S^{\rm T}(T) \bar{\Gamma}_{\{\tq\tq\},\mu}^{1^{+}}(-k_{\tq\tq};T) \,\\
&\times&S(l_{\tq};T) \,
\Delta_{\{\tqu\tq\},\rho\nu}^{1^{+}}(\ell_{\tqu\tq};T), \\
\nn {\cal M}_{\mu\nu}^{22}(T)&=&t^{f_{22}} \,
\Gamma_{\{\tq\tq\},\rho}^{1^{+}}(\ell_{\tq\tq};T) S^{\rm T}(T) \bar{\Gamma}_{\{\tq\tq\},\mu}^{1^{+}}(-k_{\tq\tq};T) \, \\
&\times&
S(l_{\tq_1};T) \,
\Delta_{\{\tq\tq\},\rho\nu}^{1^{+}}(\ell_{\tq\tq};T), 
\end{eqnarray}
where $t^f$ is the flavor matrix and can be found in Appendix~\ref{App:Flavdiq}. The color-singlet bound states constructed solely from two light up/strange quarks are written in terms of the column vector:
\begin{eqnarray}
&&\nn \begin{array}{cc}
u_{\Sigma^{*+}(uus)}= \left[
\begin{array}{c}
\{\tu\tu\}\ts \\
\{\tu\ts\} \tu \\
\end{array} \right], &
\quad u_{\Xi^{*0}(uss)}=
\left[ \begin{array}{c}
\{\tu\ts\} \ts \\
\{\ts\ts\}\tu
\end{array} \right].
\end{array} 
\end{eqnarray}
while baryons made solely of two heavy charm/bottom quarks are:
\begin{eqnarray}
&&\nn \begin{array}{cc}
u_{\Omega_{\tc\tc\tb}^{*+}}= \left[
\begin{array}{c}
\{\tc\tc\}\tb \\
\{\tc\tb\}\tc \\
\end{array} \right], & \quad
u_{\Omega_{\tc\tb\tb}^{*0}}=
\left[ \begin{array}{c}
\{\tc\tb\} \tb \\
\{\tb\tb\} \tc
\end{array} \right].
\end{array} 
\end{eqnarray}
The column vectors representing singly and doubly heavy quark baryons are:
\begin{eqnarray}
\begin{array}{rl}
\vspace{3mm}
u_{\Sigma_{\tc}^{*++}{(\tu\tu\tc)}}=
\left[ \begin{array}{c}
\{\tu\tu\} \tc \\
\{\tu\tc\} \tu
\end{array} \right], & \quad u_{\Xi_{\tc\tc}^{*++}{(\tu\tc\tc)}}= \left[
\begin{array}{c}
\{\tu\tc\}\tc \\
\{\tc\tc\} \tu \\
\end{array} \right] \, , \nn \\
\vspace{3mm}
u_{\Omega_{\tc}^{*0}{(\ts\ts\tc)}}=
\left[ \begin{array}{c}
\{\ts\ts\} \tc \\
\{\ts\tc\} \ts
\end{array} \right], &\quad u_{\Omega_{\tc\tc}^{*+}{(\ts\tc\tc)}}= \left[
\begin{array}{c}
\{\ts\tc\} \tc \\
\{\tc\tc\} \ts \\
\end{array} \right] , \\
\vspace{3mm}
u_{\Sigma_{\tb}^{*+}{(\tu\tu\tb)}}=
\left[ \begin{array}{c}
\{\tu\tu\} \tb \\
\{\tu\tb\} \tu
\end{array} \right], & \quad u_{\Xi_{\tb\tb}^{0*}{(\tu\tb\tb)}}= \left[
\begin{array}{c}
\{\tu\tb\}\tb \\
\{\tb\tb\} \tu \\
\end{array} \right], \\
u_{\Omega_{\tb}^{*-}{(\ts\ts\tb)}}=
\left[ \begin{array}{c}
\{\ts\ts\} \tb \\
\{\ts\tb\}\ts
\end{array} \right], & \quad u_{\Omega_{\tb\tb}^{*-}{(\ts\tb\tb)}}= \left[
\begin{array}{c}
\{\ts\tb\}\tb \\
\{\tb\tb\}\ts \\
\end{array} \right] ~.
\end{array}
\end{eqnarray}
%
%
%
%
%
%
%
%
The screening mass of the baryon under study, denoted by $M_B$, is determined by requiring that the eigenvalue of Eq.~(\ref{eq:D2}) be equal to unity at $P^2=-M_B^2$.
Having introduced the ingredients necessary to describe the positive parity baryons, we now proceed to the analysis of their parity partners in the following subsection.
\subsection{ Parity Partners: Spin-$3/2^-$ Baryons}
\label{secc-n}
We now turn to the negative parity sector. In the limit of exact chiral symmetry, positive and negative parity states would be degenerate. The substantial mass splittings observed in the hadron spectrum at $T=0$ MeV therefore provide a direct manifestation of dynamical chiral symmetry breaking. Representative examples include the $\rho_{1^-}(770\,\mathrm{MeV})$ and ${a}_{1^+}(1260\,\mathrm{MeV})$ mesons, as well as the $\Delta_{3/2^+}(1232\,\mathrm{MeV})$ and $\Delta_{3/2^-}(1700\,\mathrm{MeV})$ baryons, whose masses differ by approximately $500\,\mathrm{MeV}$. A similar pattern is observed in the nucleon sector, where the $N_{1/2^+}(939\,\mathrm{MeV})$ and $N^{*}_{1/2^-}(1535\,\mathrm{MeV})$ states are separated by about $600\,\mathrm{MeV}$.\\
Within the BSE and FE framework, the mass splittings between positive and
negative parity states are associated with dynamical chiral symmetry
breaking and with the different Dirac structures of their corresponding
amplitudes. Consequently, the thermal evolution of spin-$3/2$ parity
partners provides a sensitive probe of the progressive restoration of
chiral symmetry.\\
In the present symmetry preserving CI treated in the rainbow-ladder truncation, the construction of the negative parity spin-$3/2$ baryons requires special care because it is constrained by the flavor structure of the underlying diquark correlations.\\
Since the flavor wave functions of spin-$3/2$ baryons are completely symmetric, the corresponding diquark correlations must also belong to the symmetric flavor representation~\cite{Yin:2021uom}. Negative  parity V diquarks, however, belong only to the antisymmetric flavor representation (flavor antitriplet), whereas AV diquarks belong to the symmetric flavor representation (flavor sextet)~\cite{Chen:2024emt}. Consequently, the flavor symmetric negative parity diquark correlations required for spin-$3/2$ baryons such as $\{uu\}_{1^-}$, $\{ss\}_{1^-}$, $\{cc\}_{1^-}$, and $\{bb\}_{1^-}$ are not supported within this framework. Therefore, following Ref.~\cite{Yin:2021uom}, the negative parity spin-$3/2$ baryons are constructed using the same positive parity AV diquark correlations employed in the corresponding positive parity states. The distinction between the two parity sectors is introduced through the
Dirac structure of the quark-diquark FA, with $G^{+}=\mathbb{I}_{\rm D}$ and
$G^{-}=\gamma_5$, as defined in Eq.~\eqref{base}, following the
prescription given in Appendix~B of Ref.~\cite{Yin:2021uom}. All other
aspects of the calculation are implemented analogously to the procedure
described in Sec.~\ref{secc-p}.
It is worth recalling that the AV diquark correlations employed in this work are those obtained from the rainbow-ladder quark-quark BSE. As discussed in Refs.~\cite{Roberts:2011wy,Chen:2012qr,Roberts:2011cf}, these correlations arise as poles of the quark-quark scattering matrix within the rainbow-ladder truncation and should be interpreted as dynamical correlations rather than physical asymptotic diquark states.\\
Having established the construction of the negative parity states within
the present framework, we now proceed to discuss their screening masses
and compare their thermal evolution with that of their positive parity
partners, paying particular attention to the temperature evolution of
the parity splitting across the different flavor sectors.
\section{Results}
\label{Results}
\begin{figure}[t]
    \centering
    \includegraphics[width=1\linewidth]{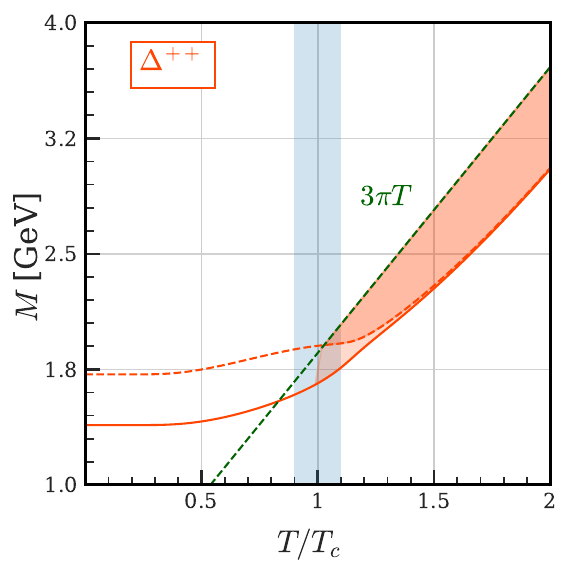}
    \vspace{-0.7cm}
    \caption{\justifying Screening masses for the $\Delta^{++}$ baryon and its negative parity partner. The solid and dashed lines correspond to the positive and negative parity states, respectively. The shaded blue band indicates the crossover region, while the green dashed line represents the asymptotic free theory limit, $3\pi T$. The shaded orange area between the free theory limit and the baryon screening masses highlights the difference between the interacting states and the asymptotic behavior.}
    \label{fig:uuu-p}
\end{figure}

\begin{figure*}[htbp]
    \centering
\includegraphics[width=1\linewidth]{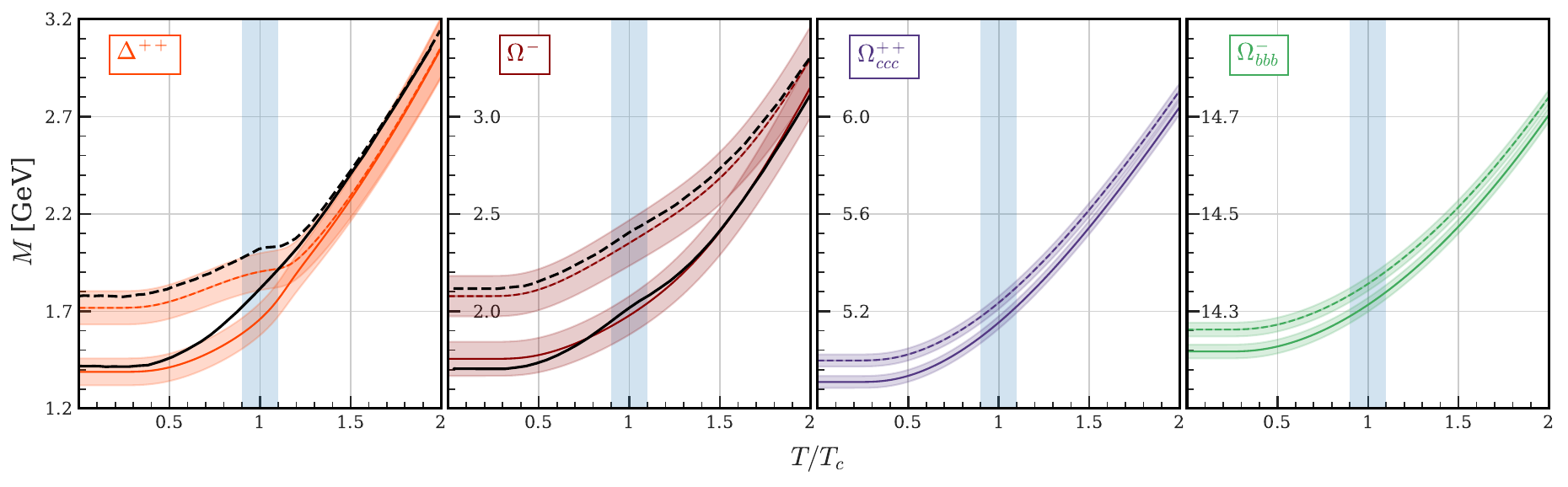}
    \vspace{-5mm}
    \caption{\justifying Screening masses of spin-3/2 baryons composed of three identical quarks ($qqq$) as functions of temperature. The first two panels show the lightest baryons composed of three identical quarks, $\Delta^{++}$ and $\Omega^{-}$, whereas the last two panels display the fully-heavy $\Omega^{++}_{\tc\tc\tc}$ and $\Omega^{-}_{\tb\tb\tb}$ systems. Solid and dashed curves correspond to the positive and negative parity screening masses, respectively. The shaded blue band indicates the crossover region. The shaded bands indicate estimated uncertainties of $5\%$ for the $\Delta^{++}$ and $\Omega^{-}$, $0.5\%$ for the $\Omega^{++}_{\tc\tc\tc}$, and $0.1\%$ for the $\Omega^{-}_{\tb\tb\tb}$. For the light-sector baryons, the screening masses obtained in the present work are also compared with the corresponding results of Ref.~\cite{Chen:2024emt}, shown by the black curves.
} 
    \label{triply-LL}
\end{figure*}

This section presents our results for the screening masses of spin-$3/2$ baryons, with particular emphasis on the comparison between positive and negative parity states. The analysis is based on the finite-temperature diquark masses and BSAs reported in Ref.~\cite{Gutierrez-Guerrero:2026hhu}; the corresponding $T=0$ MeV values are given in Table~\ref{fbodt0}.\\
\begin{table}[b]
\centering
\caption{\justifying
\label{fbodt0}
Masses (in GeV) at $T=0$ MeV of the AV diquarks, obtained using the parameters listed in Tables~\ref{parameters} and~\ref{table-M}. Further details regarding the calculation of these masses within the CI model can be found in Ref.~\cite{Gutierrez-Guerrero:2026hhu}.}
\begin{tabular}{@{\extracolsep{0.3 cm}} cc|cc}
\hline
\hline
Diquark & Mass & Diquark & Mass \\
\hline
$\{ud\}$ & 1.05 & $\{ub\}$ & 5.38 \\
 \rule{0ex}{2.5ex}
$\{us\}$ & 1.15 & $\{sb\}$ & 5.47 \\
 \rule{0ex}{2.5ex}
$\{ss\}$ & 1.25 & $\{cb\}$ & 6.33 \\
 \rule{0ex}{2.5ex}
$\{uc\}$ & 2.14 & $\{cc\}$ & 3.19 \\
 \rule{0ex}{2.5ex}
$\{sc\}$ & 2.24 & $\{bb\}$ & 9.43 \\
\hline
\hline
\end{tabular}
\end{table}
We first consider baryons composed of three identical quarks, for which the calculation of the screening masses is comparatively simpler since only a single diquark correlation contributes to the Faddeev kernel. The lightest state in this sector is the $\Delta^{++}$ baryon, composed exclusively of $\tu$ quarks. The temperature dependence of its positive and negative parity screening masses is presented in Fig.~\ref{fig:uuu-p}, together with the asymptotic free theory limit, $3\pi T$.
At low temperatures, both parity partners remain close to their respective $T=0$ MeV screening masses, indicating that thermal effects are still relatively weak in this regime.
 As the temperature approaches the critical region, indicated by the shaded blue band, the positive parity state begins to increase, whereas the negative parity baryon exhibits a slight reduction. At higher temperatures, the two states progressively converge; within the present calculation, the percentage difference between the positive and negative parity states is reduced to approximately $0.33\%$ for this baryon. Such behavior is commonly associated with chiral symmetry restoration. Moreover, Fig.~\ref{fig:uuu-p} shows that, at sufficiently high temperatures, the screening masses increase approximately linearly with $T$ and approach the asymptotic free theory limit represented by the $3\pi T$ line. 
In particular, at $T=2\,T_c$, the result obtained in the present work differs from the free-theory limit by approximately $17.9\%$.\\
The remaining baryons composed of three identical quarks, namely the $\Omega^{-}$, $\Omega_{\tc\tc\tc}^{++}$, and $\Omega_{\tb\tb\tb}^{-}$, are displayed in Fig.~\ref{triply-LL}. 
 The first two panels correspond to the light sector, namely the $\Delta^{++}$ and $\Omega^{-}$ baryons, whereas the last two panels display the triply heavy states $\Omega_{\tc\tc\tc}^{++}$ and $\Omega_{\tb\tb\tb}^{-}$. In each case, the positive parity states (solid lines) are compared with their corresponding negative parity partners (dashed lines).
 The screening masses at $T=0$ MeV, listed in Table~\ref{tab:screening_25}, are in good overall agreement with the corresponding results reported in Ref.~\cite{Yin:2021uom}. For all the baryons considered, the relative differences remain below $10\%$. In the light sector, the $\Delta^{++}$ and $\Omega^{-}$ screening masses differ by approximately $7$-$10\%$, depending on the parity assignment. By contrast, the agreement improves significantly in the fully-heavy sector, where the corresponding differences are only about $0.2\%$ for the $\Omega_{ccc}^{++}$ and $1.6\%$ for the $\Omega_{bbb}^{-}$, independently of the parity. It is worth noting that the CI framework has been shown to reproduce the experimental baryon spectrum reported by the Particle Data Group~\cite{ParticleDataGroup:2026aaa} with typical deviations of only a few percent over a broad range of light, heavy-light, and heavy baryons. Moreover, for states that have not yet been experimentally observed, the corresponding CI predictions are generally consistent with those obtained using other theoretical approaches~\cite{Qin:2019hgk,Bahtiyar:2020uuj,Li:2022vbc}. This overall agreement provides additional confidence in the zero-temperature spectrum employed as the baseline for the present finite-temperature analysis.
The table also presents the screening masses at $T=2\,T_c$, together with the relative percentage difference between the positive and negative parity screening masses,
\begin{equation}
\Delta_T=
\frac{\left|M_{3/2^-}(T)-M_{3/2^+}(T)\right|}
{M_{3/2^-}(T)}
\times100,
\label{eq:delta_percentage}
\end{equation}
where $M_{3/2^+}(T)$ and $M_{3/2^-}(T)$ denote the positive and negative parity screening masses, respectively. The quantities labeled $\Delta_{T_0}$, $\Delta_{T_1}$, and $\Delta_{T_2}$ correspond to $\Delta_T$ evaluated at $T=0$ MeV, $T=2\,T_c$, and $T=0.3$ GeV, respectively.
For example, in the $\Delta^{++}$ channel, the percentage difference 
decreases markedly from $19.19\%$ at $T=0$ MeV to only $0.33\%$ at 
$T=2\,T_c$, while for the $\Omega^{-}$ it decreases from $15.38\%$ 
to $4.55\%$. These results indicate a substantial reduction of the 
parity splitting with increasing temperature in the light sector. 
In contrast, this reduction becomes considerably less pronounced as 
the heavy-quark content increases: for the $\Omega_{ccc}^{++}$, 
$\Delta_T$ decreases only from $1.80\%$ to $1.15\%$, whereas for the 
$\Omega_{bbb}^{-}$ it remains nearly unchanged, decreasing from 
$0.28\%$ to $0.27\%$. Thus, although the percentage difference between 
parity partners decreases in all the baryons listed in Table~\ref{tab:screening_25}, 
the effect is much more pronounced for the light baryons and becomes 
progressively weaker in the fully-heavy sector.\\
\begin{table}[b]
\centering
\caption{\justifying
\label{screening_25}
Screening masses (in GeV) at $T=0$ MeV and $T=2\,T_c$ for the baryons shown in Fig.~\ref{triply-LL}. The last two columns report the relative percentage differences, $\Delta_T$, between the positive ($J^P=3/2^+$) and negative parity ($J^P=3/2^-$) screening masses, as defined in Eq.~(\ref{eq:delta_percentage}). The entries labeled $\Delta_{T_0}$ and $\Delta_{T_1}$ correspond to $\Delta_T$ evaluated at $T=0$ MeV and $T=2\,T_c$, respectively.
}
\begin{tabular}{@{\extracolsep{0.15cm}}c|c|c|c|c|c}
\hline
\rule{0ex}{2.5ex}
Baryon & $J^P$ & $T=0$ MeV & $T=2\,T_c$ & $\Delta_{T_0}$ &  $\Delta_{T_1}$\\
\hline
\rule{0ex}{2.5ex}
\multirow{2}{*}{$\Delta^{++}$}
& $3/2^{+}$ & 1.39 & 3.05 &
\multirow{2}{*}{19.19} &
\multirow{2}{*}{0.33} \\
\rule{0ex}{2.5ex}
& $3/2^{-}$ &
1.72 &
3.06 & & \\
\hline
\rule{0ex}{2.5ex}
\multirow{2}{*}{$\Omega^-$}
& $3/2^{+}$ & 1.76 & 3.15 &
\multirow{2}{*}{15.38} &
\multirow{2}{*}{4.55} \\
\rule{0ex}{2.5ex}
& $3/2^{-}$ &
2.08 &
3.30 & & \\
\hline
\rule{0ex}{2.5ex}
\multirow{2}{*}{$\Omega_{\tc\tc\tc}^{++}$}
& $3/2^{+}$ & 4.91 & 6.04 &
\multirow{2}{*}{1.80} &
\multirow{2}{*}{1.15} \\
\rule{0ex}{2.5ex}
& $3/2^{-}$ &
5.00 &
6.11 & & \\
\hline
\rule{0ex}{2.5ex}
\multirow{2}{*}{$\Omega_{\tb\tb\tb}^-$}
& $3/2^{+}$ & 14.22 & 14.70 &
\multirow{2}{*}{0.28} &
\multirow{2}{*}{0.27} \\
\rule{0ex}{2.5ex}
& $3/2^{-}$ &
14.26 &
14.74 & & \\
\hline
\end{tabular}
\label{tab:screening_25}
\end{table}
For all baryons shown in Fig.~\ref{triply-LL}, the screening masses increase with temperature, although the thermal dependence becomes progressively weaker as the quark masses increase. In particular, the triply heavy baryons exhibit only mild thermal modifications below the crossover region, indicating a comparatively weak thermal response of these systems. By contrast, the light sector displays a stronger temperature dependence together with a tendency toward parity partner degeneracy near and above the crossover region, consistent with the gradual restoration of chiral symmetry.\\
Comparison with the results of Ref.~\cite{Chen:2024emt} further shows that the qualitative features of the thermal evolution are robust against variations in the treatment of the Faddeev kernel. For the light sector baryons, both approaches yield very similar positive parity screening masses, whereas somewhat larger differences are observed for the negative parity states. In both parity channels, however, these differences decrease with increasing temperature.\\
\begin{figure*}[t]
    \centering
\includegraphics[width=1\linewidth]{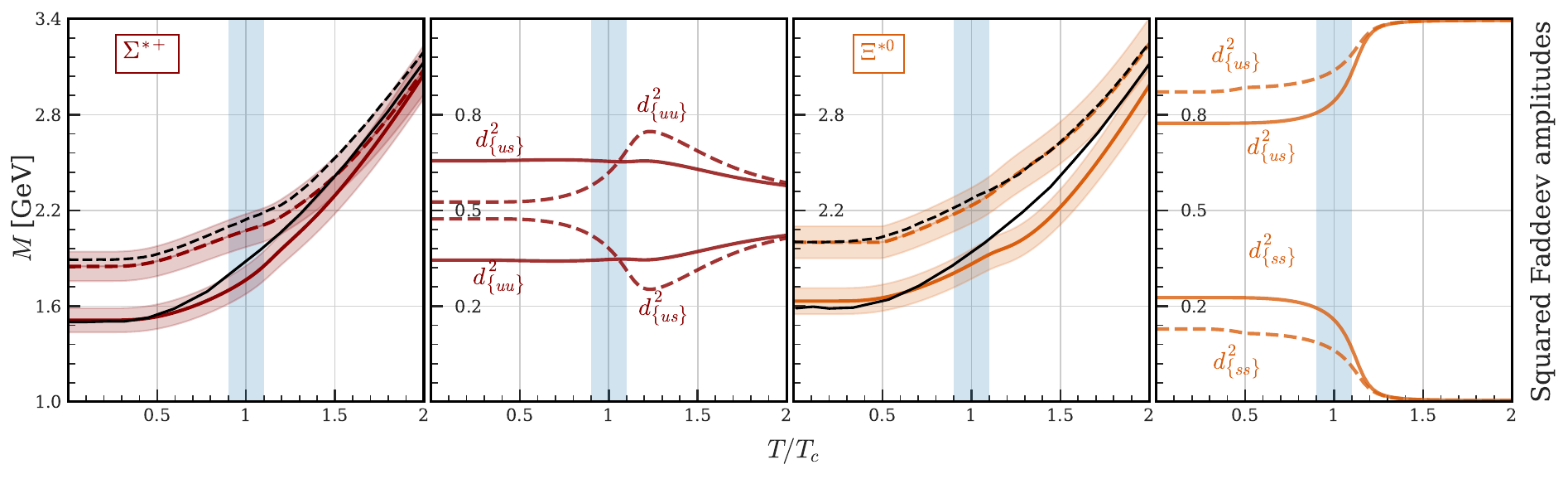}
  \caption{\label{triply-LH} \justifying
Screening masses and amplitudes of the ($\tqu\tq\tq$) spin-$3/2$ baryons composed of $\tu$ and $\ts$ quarks. The screening masses are presented in the first and third panels, while the squared Faddeev amplitudes are shown in the second and fourth panels. In each case, the positive parity states (solid lines) are compared with their respective parity partners (dashed lines). The results are computed in the temperature range from $T=0$ MeV up to $T/T_c=2$. The shaded blue band indicates the crossover region. Uncertainty bands of 5\% are included. For comparison, the corresponding screening masses from Ref.~\cite{Chen:2024emt} are shown by the black curves.}
\end{figure*}
As in the meson sector, we define a characteristic temperature, $T_H$, above which the baryon begins to exhibit a qualitative change in its thermal behavior. In this context, $T_H$ is defined as:
\bea T_{\chi}=\min\left\{T \,\Big|\, \|\nabla M\| > \chi\,\|\nabla M\|_{\max}\right\},\label{theq}\eea
where $\chi=0.05$ and $\chi=0.5$ correspond to $T_H$ and $T_{HC}$, respectively. Here, $\|\nabla M\|$ denotes the thermal variation of the screening mass, and $\|\nabla M\|_{\max}$ its maximum value in the explored temperature range. The temperature $T_H$ marks the onset of noticeable thermal modifications, while $T_{HC}$ identifies the region where the thermal variation becomes large, typically close to the critical temperature.
For comparison, lattice QCD calculations performed by the HotQCD Collaboration estimate the pseudocritical temperature associated with the chiral crossover to be
$T_c \approx 154 \pm 9~\text{MeV}$
for QCD with physical quark masses~\cite{Bazavov:2011nk}.
Previous studies within the same  CI framework have reported the critical temperature $T_c = 197~\mathrm{MeV}$,
as discussed in Ref.~\cite{Chen:2024emt}. In that study, the critical temperature is identified from the merging of the negative Nambu and Wigner solutions of the gap equation at finite-temperature, signaling chiral symmetry restoration and deconfinement within the model. Although this value is larger than the pseudocritical temperature typically reported by lattice-QCD simulations, it is consistent with previous rainbow-ladder CI studies once normalized with the corresponding V mass scale.\\
The values of the characteristic temperatures $T_H$ and $T_{HC}$ for the baryons displayed in Fig.~\ref{triply-LL} are listed in Table~\ref{triply-TH}.
\begin{table}[t]
\centering
\caption{\justifying Values of the characteristic temperature $T_H$ and $T_{HC}$ in GeV for the positive and negative parity triply flavored spin-$3/2$ baryons displayed in Fig.~\ref{triply-LL}.}
\begin{tabular}{@{\extracolsep{0.1cm}}c|cc|cc|cc|cc}
\hline
& \multicolumn{2}{c|}{$\Delta^{++}$}
& \multicolumn{2}{c|}{$\Omega^-$}
& \multicolumn{2}{c|}{$\Omega_{\tc\tc\tc}^{++}$}
& \multicolumn{2}{c}{$\Omega_{\tb\tb\tb}^-$} \\
\hline
\rule{0ex}{2.5ex}
& $3/2^{+}$ & $3/2^{-}$
& $3/2^{+}$ & $3/2^{-}$
& $3/2^{+}$ & $3/2^{-}$
& $3/2^{+}$ & $3/2^{-}$ \\
\hline
\rule{0ex}{2.5ex}
$T_H$
& 0.08 & 0.07
& 0.08 & 0.07
& 0.07 & 0.07
& 0.08 & 0.08 \\
\hline
\rule{0ex}{2.5ex}
$T_{HC}$
& 0.20 & 0.25
& 0.25 & 0.28
& 0.17 &0.18
& 0.21 & 0.21 \\
\hline
\end{tabular}
\label{triply-TH}
\end{table}
Unlike the nearly uniform behavior observed for $T_H$, the values of $T_{HC}$ exhibit a clearer dependence on the parity channel. For all the baryons considered in Table~VI, one finds
\begin{equation}
T_{HC}^{3/2^-} \geq T_{HC}^{3/2^+},
\label{rel_thc}
\end{equation}
indicating that the onset of strong thermal modifications occurs at comparable or slightly higher temperatures for the negative parity states than for their positive parity counterparts. This parity dependence is most pronounced in the light sector and gradually weakens as the quark mass increases, becoming negligible for the triply bottom baryon, for which both parity states yield the same value of $T_{HC}$.
A useful comparison can be made with the characteristic degeneracy temperature introduced in Ref.~\cite{Chen:2024emt}, where the reduction of the parity splitting was quantified through the temperature $T_c^{\delta m}$, defined as the point at which the relative mass splitting between parity partners decreases to $0.5\%$.
For the spin-$3/2$ decuplet baryons, they obtained
$T_{c}^{\delta m,\Delta^\pm}=1.47\,T_c$,
$T_{c}^{\delta m,\Sigma^{*\pm}}=2.71\,T_c$,
$T_{c}^{\delta m,\Xi^{*\pm}}=3.64\,T_c$, and
$T_{c}^{\delta m,\Omega^\pm}=4.42\,T_c$.
These values reveal a clear flavor dependence, indicating that the temperature required to achieve near degeneracy increases systematically with the strangeness content of the baryon.
By contrast, the characteristic temperatures $T_H$ and $T_{HC}$ introduced in the present work probe earlier stages of the thermal evolution, namely the onset of visible thermal modifications and the emergence of strong medium effects, respectively. For the $\Delta^{++}$ baryon, we obtain $T_H=(0.08,\,0.07)$ GeV and $T_{HC}=({0.20},\,0.25)$ GeV for the positive and negative parity states, respectively. These values indicate that noticeable thermal effects arise well before the temperature at which the parity partners become nearly degenerate. Taken together, the results show that $T_H$, $T_{HC}$, and $T_c^{\delta m}$ characterize successive stages of the thermal evolution: the onset of visible thermal modifications, the emergence of strong medium effects, and, finally, parity doubling at considerably higher temperatures.\\
%
Baryons composed of three light quarks, two of which carry distinct flavors, are displayed in Fig.~\ref{triply-LH}.
 In both the $\Sigma^{*+}$ and $\Xi^{*0}$ channels, the positive parity states increase monotonically with temperature, while their negative parity partners exhibit a much softer thermal dependence below the crossover region. As the temperature increases, the masses of the parity partners progressively approach each other, leading to a substantial reduction of the parity splitting at high temperature.
The squared Faddeev amplitudes shown in the second and fourth panels 
remain nearly constant at low temperatures. As the temperature approaches 
and exceeds the crossover region, their evolution becomes more pronounced, 
although the magnitude and pattern of the reorganization depend on the 
baryon channel.\\
The screening masses of the positive and negative parity baryons at $T=0$ MeV and $T=2\,T_c$, together with the characteristic temperatures $T_H$ and $T_{HC}$, are summarized in Table~\ref{table-us}. A notable feature is the strong reduction of the parity splitting at $T=2\,T_c$: the relative difference between parity partners decreases to only $0.97\%$ in the $\Sigma^{*+}$ channel and $8.02\%$ in the $\Xi^{*0}$ system. This behavior is consistent with the onset of parity doubling expected as chiral symmetry is progressively restored.
\begin{table}[t]
\centering
\caption{\justifying
Screening masses (in GeV) at $T=0$ MeV and $T=2\,T_c$ for the positive and negative parity spin-$3/2$ baryons composed of $\tu$ and $\ts$ quarks. The columns labeled $\Delta_{T_0}$ and $\Delta_{T_1}$ report the values of $\Delta_T$ defined in Eq.~(\ref{eq:delta_percentage}). The fifth and sixth columns list the corresponding values of $T_H$ and $T_{HC}$ in GeV.}
\begin{tabular}{c|c|c|c|c|c|c|c}
\hline
Baryon & State & $T=0$ MeV& $T=2\,T_c$ & $T_H$ & $T_{HC}$ & $\Delta_{T_0}$ & $\Delta_{T_1}$\\
\hline
\rule{0ex}{2.5ex}
\multirow{2}{*}{$\Sigma^{\ast +}$}
& $3/2^{+}$ & 1.51 & 3.05 & 0.08 & 0.21 & \multirow{2}{*}{18.38} & \multirow{2}{*}{0.97} \\
\rule{0ex}{2.5ex}
& $3/2^{-}$ & 1.85 & 3.08 & 0.07 & 0.26 & & \\
\hline
\rule{0ex}{2.5ex}
\multirow{2}{*}{$\Xi^{\ast 0}$}
& $3/2^{+}$ & 1.63 & 2.98 & 0.08 & 0.27 & \multirow{2}{*}{18.5} & \multirow{2}{*}{8.02} \\
\rule{0ex}{2.5ex}
& $3/2^{-}$ & 2.00 & 3.24 & 0.10 & 0.22  & & \\
\hline
\end{tabular}
\label{table-us}
\end{table}
The values of $T_{HC}$ exhibit a clear dependence on the parity channel. 
For the $\Sigma^{*+}$ baryon, the negative parity state satisfies the 
relation in Eq.~(\ref{rel_thc}), with $T_{HC}^{3/2^-}=0.26$ GeV compared with $T_{HC}^{3/2^+}=0.21$ GeV. In contrast, the $\Xi^{*0}$ baryon exhibits the opposite behavior, with $T_{HC}^{3/2^-}=0.22$ GeV and $T_{HC}^{3/2^+}=0.27$ GeV.\\

For the light baryons considered in this sector, an important feature
of the present approach is that, as the temperature increases, the
screening masses of the parity partners progressively approach each other.
At the same time, their corresponding Faddeev amplitudes undergo a temperature-dependent reorganization, which does not necessarily lead to similar
quark--diquark compositions in the two parity channels.
As shown in Fig.~8, for the $\Xi^{\ast 0}$ baryon,
the $\{us\}$ diquark correlation becomes dominant at high
temperatures in both parity channels, whereas the $\{ss\}$
contribution is strongly suppressed. In contrast, for the
$\Sigma^{\ast +}$ baryon, the dominant diquark correlations
remain different in the positive and negative parity channels,
even as their screening masses approach degeneracy.
This behavior is consistent with the fact
that the convergence of the screening masses does not
require identical Faddeev eigenvectors. Indeed, the
screening masses are obtained from the condition that
the corresponding Faddeev eigenvalue reaches unity,
whereas the relative contributions of the quark--diquark
configurations are determined by the associated eigenvectors.
Consequently, different internal compositions may persist
even when the screening masses of the parity partners
become nearly degenerate.

A temperature-dependent redistribution of
the diquark contributions has also been reported in
Ref.~\cite{Chen:2024emt}.
In particular, for the
negative parity $\Sigma^{\ast}$ baryon, the dominant
diquark contribution changes with increasing temperature,
with the $\{us\}$ correlation becoming dominant at
high temperatures. This illustrates that the relative
importance of different quark--diquark configurations
can change significantly with temperature.

Although the detailed hierarchy of the
Faddeev-amplitude components differs between the two
calculations, both exhibit a nontrivial thermal
reorganization of the diquark contributions.
These differences may be associated with
the distinct treatments of the Faddeev equation employed
in the two studies.
In the present work, the baryon amplitudes are obtained
within the static approximation, whereas
Ref.~\cite{Chen:2024emt} employs the momentum-fraction
approximation introduced in Ref.~\cite{Wang:2013wk},
which preserves a different realization of the
quark-exchange dynamics.
Nevertheless, within the present approximation,
the dynamical origin of the strong suppression of particular
Faddeev-amplitude components cannot be unambiguously
established. Therefore, the approach toward screening-mass
degeneracy should not be interpreted as a complete
convergence of the internal quark--diquark structure
of the corresponding parity partners.

\begin{figure}[t]
    \centering
    \includegraphics[width=1\linewidth]{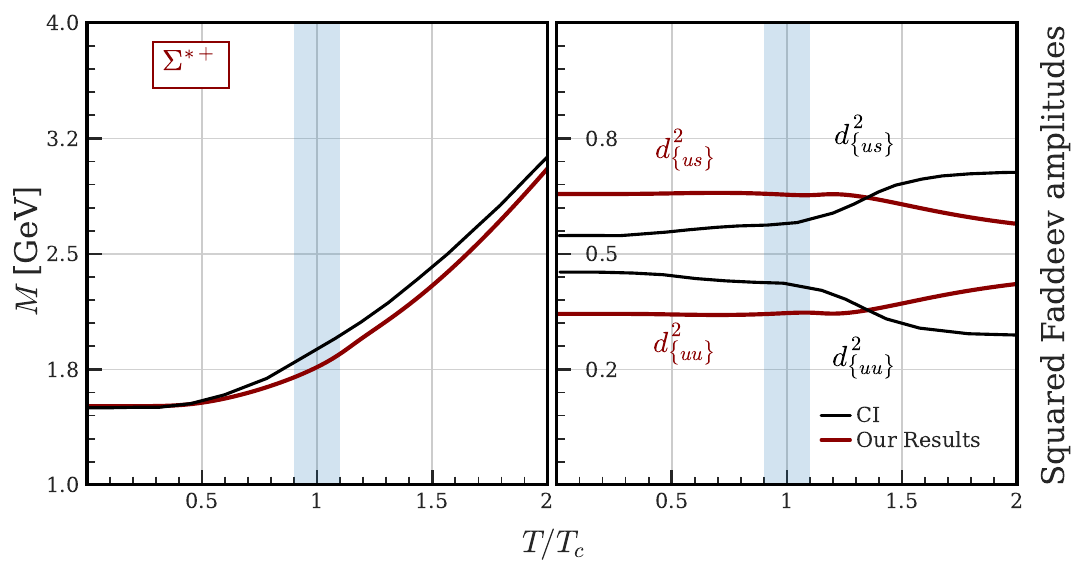}
    \caption{\justifying
    Temperature dependence of the screening masses and squared
    Faddeev amplitudes of the positive parity
    $\Sigma^{\ast +}$ baryon.
    The shaded blue band indicates the crossover region.
    The results obtained in the present work are compared
    with the corresponding predictions of
    Ref.~\cite{Chen:2024emt} within the same CI framework.}
    \label{amp-LH}
\end{figure}

To illustrate these differences explicitly, Fig.~\ref{amp-LH}
displays the squared Faddeev amplitudes for the
$\Sigma^{\ast +}$ baryon as functions of temperature,
compared with the corresponding results reported in
Ref.~\cite{Chen:2024emt}. Although the thermal evolution
of the Faddeev amplitudes differs significantly in the
two approaches, the resulting screening masses remain
remarkably similar over the entire temperature range
considered. This observation suggests that,
within the approximations considered,
the screening masses are relatively insensitive to the
detailed structure of the Faddeev amplitudes,
providing additional evidence for the robustness of the
predicted thermal evolution.

In Table~\ref{tab:diquark_amplitudes}, we present the
squared components of the Faddeev amplitude
obtained for the positive parity
$\Sigma^{\ast +}$ baryon at $T=0$ MeV.
While the relative weights of the individual diquark
components differ somewhat from those reported in
Ref.~\cite{Chen:2024emt}, both studies identify the
$\{\tu\ts\}$ correlation as the dominant contribution
to the $\Sigma^{\ast +}$ structure. Remarkably, these differences translate into only negligible changes in the screening masses. For example, the screening mass of the $\Sigma^{\ast +}$ baryon at $T=0$~MeV is $1.50~\mathrm{GeV}$ in Ref.~\cite{Chen:2024emt}, compared with $1.51~\mathrm{GeV}$ obtained in the present work. This observation indicates that the predicted screening masses are robust against moderate variations in the detailed structure of the Faddeev amplitudes. \\

\begin{table}[htbp]
\centering
\caption{\justifying
Squared components of the Faddeev amplitude
for the positive parity $\Sigma^{*+}$ baryon
at $T=0$ MeV in the present work compared with the
corresponding CI values used in Ref.~\cite{Chen:2024emt}.}
\begin{tabular}{@{\extracolsep{0.5cm}}c c c}
\hline\hline
 & This work & CI in Ref.~\cite{Chen:2024emt} \\
\hline
$d_{\{uu\}}^2$ & $0.37$ & $0.46$ \\
$d_{\{us\}}^2$ & $0.63$ & $0.54$ \\
\hline\hline
\end{tabular}
\label{tab:diquark_amplitudes}
\end{table}
%


The light sector therefore serves as a benchmark for the thermal behavior
of spin-$3/2$ baryons, exhibiting an overall increase of the screening
masses with temperature, a progressive reduction of the separation between
parity partners, and the general trend summarized in Eq.~(\ref{rel_thc}).

\begin{figure*}[htbp]
    \centering
    \includegraphics[width=1\linewidth]{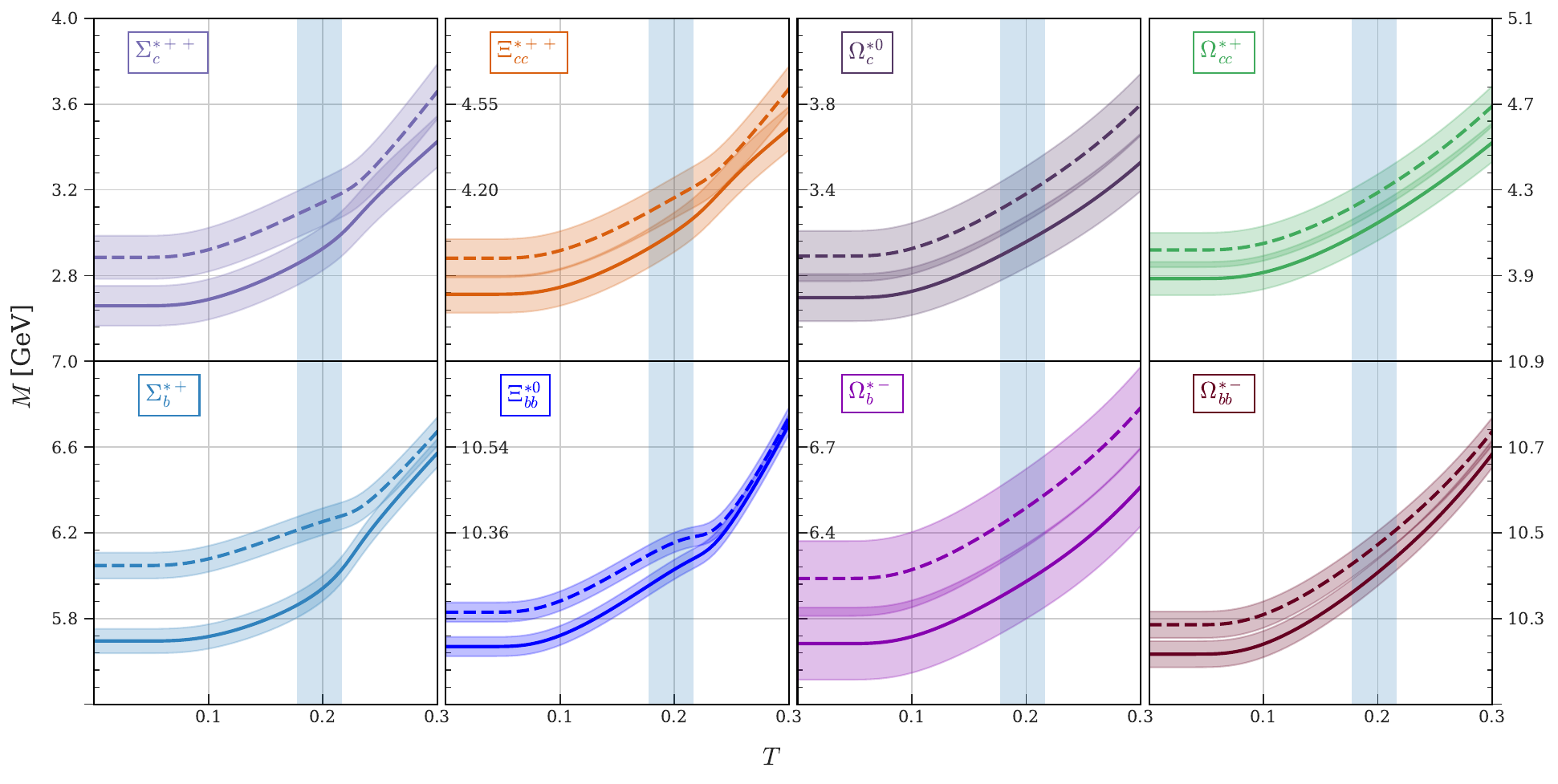}
\caption{\label{udosq}\justifying Masses of the ($q_1qq$) spin-3/2 baryons composed of heavy-light quarks. Both masses and temperatures are expressed in GeV. In each case, the positive parity states (solid lines) are compared with their respective parity partners (dashed lines). The shaded blue band indicates the crossover region. Uncertainty bands of 3.5\%, 2.0\%, 3.8\%, 2.0\%, 1.0\%, 0.2\%, 2.1\%, and 0.3\% for $\Sigma^{*++}_{\tc}$, $\Xi^{*++}_{\tc\tc}$, $\Omega^{*0}_{\tc}$, $\Omega^{*+}_{\tc\tc}$, $\Sigma^{*+}_{\tb}$, $\Xi^{*0}_{\tb\tb}$, $\Omega^{*-}_{\tb}$, and $\Omega^{*-}_{\tb\tb}$, respectively, are included.}
\end{figure*}

We now turn to baryons containing one or two charm and/or bottom quarks, whose screening masses are presented in Fig.~\ref{udosq}.
\begin{figure*}[htbp]
    \centering
      \includegraphics[width=1.0\linewidth]{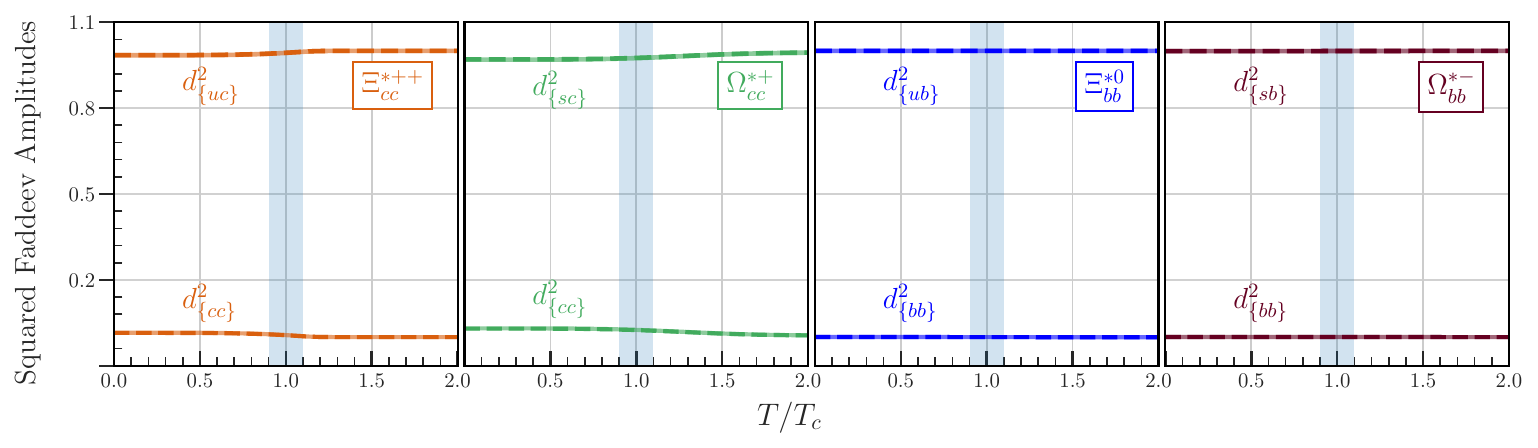}
\caption{\label{udosq3}\justifying 
Thermal evolution of the squared Faddeev amplitudes (FAs) for baryons containing two heavy quarks. From left to right, the panels correspond to the $\Xi_{\tc\tc}^{\ast ++}$, $\Omega_{\tc\tc}^{\ast +}$, $\Xi_{\tb\tb}^{\ast 0}$, and $\Omega_{\tb\tb}^{\ast -}$ baryons. Solid and dashed lines denote the positive and negative parity states, respectively. In each case, the dominant and subleading diquark components are displayed as functions of $T/T_c$. The shaded blue band indicates the crossover region. In all cases, the dominant diquark component remains largely unchanged with temperature and continues to dominate the structure of the corresponding baryon throughout the entire temperature range considered.
}
\end{figure*}
The positive parity states remain nearly constant at low temperatures 
and then increase steadily as the temperature approaches and exceeds 
the crossover region. In contrast, their negative parity partners 
generally exhibit a weaker thermal dependence below the crossover 
region. As the temperature increases, the relative separation between 
the parity partners evolves differently depending on the flavor 
content, indicating that the thermal evolution of the parity splitting 
is not uniform across the heavy-baryon sector.\\
A clear flavor dependence is also observed. The charm-containing 
baryons, such as the $\Sigma_{c}^{*++}$ and $\Omega_{cc}^{*+}$, 
experience more pronounced thermal modifications, whereas the 
doubly-bottom states remain comparatively stable over a wider 
temperature range. In particular, the $\Xi_{bb}^{*0}$ and 
$\Omega_{bb}^{*-}$ baryons exhibit only mild thermal variations below 
the crossover region, reflecting the enhanced thermal stability 
associated with the increasing heavy-quark content.\\
Above the crossover region, the screening masses of all states increase 
more rapidly with temperature, although the evolution of the relative 
separation between parity partners remains flavor dependent. The 
positive and negative parity heavy spin-$3/2$ screening masses at 
$T=0$~MeV and $T=0.3$ GeV, together with the corresponding relative 
percentage differences and the characteristic temperatures $T_H$ and 
$T_{HC}$, are summarized in Table~\ref{mspyn}.\\
\begin{table*}[htbp]
\centering
\caption{\justifying
Screening masses (in GeV) of spin-3/2 heavy baryons in the CI for both positive and negative parity states at $T=0$ MeV and $T=0.3$ GeV. The rows labeled $\Delta_{T_0}$ and $\Delta_{T_2}$ report the values of $\Delta_T$ in Eq.~(\ref{eq:delta_percentage}). The last two rows list the corresponding values of $T_H$ and $T_{HC}$ in GeV.}
\begin{tabular}{@{\extracolsep{0.1cm}}c|cc|cc|cc|cc|cc|cc|cc|cc}
\hline
& \multicolumn{2}{c|}{$\Sigma_{\tc}^{*++}$}
& \multicolumn{2}{c|}{$\Xi_{\tc\tc}^{*++}$}
& \multicolumn{2}{c|}{$\Omega_{\tc}^{*0}$}
& \multicolumn{2}{c|}{$\Omega_{\tc\tc}^{*+}$}
& \multicolumn{2}{c|}{$\Sigma_{\tb}^{*+}$}
& \multicolumn{2}{c|}{$\Xi_{\tb\tb}^{*0}$}
& \multicolumn{2}{c|}{$\Omega_{\tb}^{*-}$}
& \multicolumn{2}{c}{$\Omega_{\tb\tb}^{*-}$} \\
\hline
\rule{0ex}{2.5ex}
& $3/2^+$ & $3/2^-$
& $3/2^+$ & $3/2^-$
& $3/2^+$ & $3/2^-$
& $3/2^+$ & $3/2^-$
& $3/2^+$ & $3/2^-$
& $3/2^+$ & $3/2^-$
& $3/2^+$ & $3/2^-$
& $3/2^+$ & $3/2^-$ \\
\hline
\rule{0ex}{2.5ex}
$T=0$ MeV
& 2.66 & 2.89
& 3.77 & 3.92
& 2.90 & 3.09
& 3.89 & 4.02
& 5.69 & 6.05
& 10.12 & 10.19
& 6.01 & 6.24
& 10.22 & 10.29 \\
\rule{0ex}{2.5ex}
$\Delta_{T_0}$
& \multicolumn{2}{c|}{7.96}
& \multicolumn{2}{c|}{3.83}
& \multicolumn{2}{c|}{6.15}
& \multicolumn{2}{c|}{3.23}
& \multicolumn{2}{c|}{5.95}
& \multicolumn{2}{c|}{0.69}
& \multicolumn{2}{c|}{3.69}
& \multicolumn{2}{c}{0.68} \\
\hline
\rule{0ex}{2.5ex}
$T=0.3$ GeV
& 3.42 & 3.66
& 4.45 & 4.61
& 3.53 & 3.80
& 4.52 & 4.69
& 6.57 & 6.67
& 10.59 & 10.60
& 6.56 & 6.84
& 10.68 & 10.74 \\
\rule{0ex}{2.5ex}
$\Delta_{T_2}$
& \multicolumn{2}{c|}{6.56}
& \multicolumn{2}{c|}{3.47}
& \multicolumn{2}{c|}{7.11}
& \multicolumn{2}{c|}{3.62}
& \multicolumn{2}{c|}{1.50}
& \multicolumn{2}{c|}{0.09}
& \multicolumn{2}{c|}{4.09}
& \multicolumn{2}{c}{0.56} \\
\hline
\rule{0ex}{2.5ex}
$T_H$
& 0.07 & 0.07
& 0.07 & 0.07
& 0.07 & 0.07
& 0.07 & 0.07
& 0.08 & 0.07
& 0.07 & 0.07
& 0.07 & 0.07
& 0.07 & 0.07 \\
\hline
\rule{0ex}{2.5ex}
$T_{HC}$
& 0.19 & 0.23
& 0.17 & 0.23
& 0.23 & 0.21
& 0.16 & 0.19
& 0.20 & 0.24
& 0.24 & 0.25
& 0.21 & 0.24
& 0.22 & 0.23 \\
\hline
\end{tabular}
\label{mspyn}
\end{table*}
The values collected in Table~\ref{mspyn} show that the thermal 
evolution of the relative percentage difference between parity partners 
is not uniform across the heavy-baryon sector. For the 
$\Sigma_{c}^{*++}$ and $\Xi_{cc}^{*++}$ baryons, $\Delta_T$ decreases 
from $7.96\%$ to $6.56\%$ and from $3.83\%$ to $3.47\%$, respectively. 
A considerably stronger reduction is observed for the 
$\Sigma_{b}^{*+}$ and $\Xi_{bb}^{*0}$ channels, where $\Delta_T$ 
decreases from $5.95\%$ to $1.50\%$ and from $0.69\%$ to $0.09\%$, 
respectively. The $\Omega_{bb}^{*-}$ baryon also exhibits a slight 
reduction, from $0.68\%$ to $0.56\%$. In contrast, three channels show 
the opposite behavior: for the $\Omega_{c}^{*0}$, $\Omega_{cc}^{*+}$, 
and $\Omega_{b}^{*-}$ baryons, $\Delta_T$ increases from $6.15\%$ to 
$7.11\%$, from $3.23\%$ to $3.62\%$, and from $3.69\%$ to $4.09\%$, 
respectively. It is important to emphasize, however, that the apparent 
increase of the mass separation in some channels should be interpreted 
in light of the parameter dependence of the calculation. The central 
curves correspond to our chosen parameter set, while the shaded bands 
illustrate the range of variation obtained when the relevant model 
parameters are varied. These bands therefore provide an estimate of how 
much the predicted masses, and consequently the separation between 
parity partners, may deviate from the central values for alternative 
parameter choices. Thus, while the central results indicate that the 
relative parity splitting decreases with increasing temperature in most 
of the heavy baryons considered here, the behavior of the remaining 
channels should be understood together with the associated parametric 
uncertainty.\\
A qualitatively similar reduction of the parity splitting has been
reported for singly charmed baryons in Ref.~\cite{Suenaga:2024vwr},
where the positive and negative parity $\Sigma_c$ states progressively
approach degeneracy as the temperature increases. In that analysis, the
spin-$1/2$ and spin-$3/2$ states belonging to the heavy-quark spin
doublet are treated collectively, so that a direct state-by-state
comparison with the present $\Sigma_c^{*++}$ results is not possible.
Moreover, Ref.~\cite{Suenaga:2024vwr} considers pole masses within an
NJL-based heavy-quark effective framework, whereas the present study
addresses screening masses. Nevertheless, the reduction of the parity
splitting found here for the $\Sigma_c^{*++}$ channel is qualitatively
consistent with the tendency toward parity degeneracy reported in
Ref.~\cite{Suenaga:2024vwr}.\\
The characteristic temperatures $T_{HC}$ also display a dependence on 
the parity channel. For most of the heavy baryons considered in 
Table~\ref{mspyn}, the negative parity states satisfy the relation in 
Eq.~(\ref{rel_thc}), with the $\Omega_c^{*0}$ baryon being the only 
exception. This indicates a general tendency for the onset of strong 
thermal modifications to occur at comparable or slightly higher 
temperatures in the negative parity channel. Moreover, the relatively 
small variation of $T_{HC}$ across the different flavor sectors suggests 
that this characteristic temperature remains closely associated with 
the crossover region.\\
The thermal evolution of the squared FAs for a representative set of doubly heavy baryons is displayed in Fig.~\ref{udosq3}. These amplitudes provide information on the relative weight of the different diquark correlations in the baryon wave function. By construction, the sum of the squared amplitudes is normalized to unity. In all cases, the amplitudes remain nearly constant below the crossover region, indicating that the internal diquark structure is only weakly affected by thermal effects at low temperatures. Around the crossover region, highlighted by the shaded band, mild variations begin to emerge, particularly in the subleading components. Nevertheless, the dominant diquark correlation remains essentially unchanged and continues to govern the internal structure of the corresponding baryon throughout the entire temperature range considered.\\
Finally, Fig.~\ref{triply-HH} displays the thermal evolution of the screening masses of the fully-heavy spin-$3/2$ baryons with different flavor content, $\Omega_{\tc\tc\tb}^{*+}$ and $\Omega_{\tc\tb\tb}^{*0}$, for both parity channels. The corresponding screening masses at $T=0$ MeV and $T=2\,T_c$, 
together with the relative percentage differences between parity partners and the characteristic temperatures $T_H$ and $T_{HC}$, are summarized in Table~\ref{mspyn-heavy}.
\begin{figure}[htb]
    \centering
\hspace{-0.3cm}\includegraphics[width=1.03\linewidth]{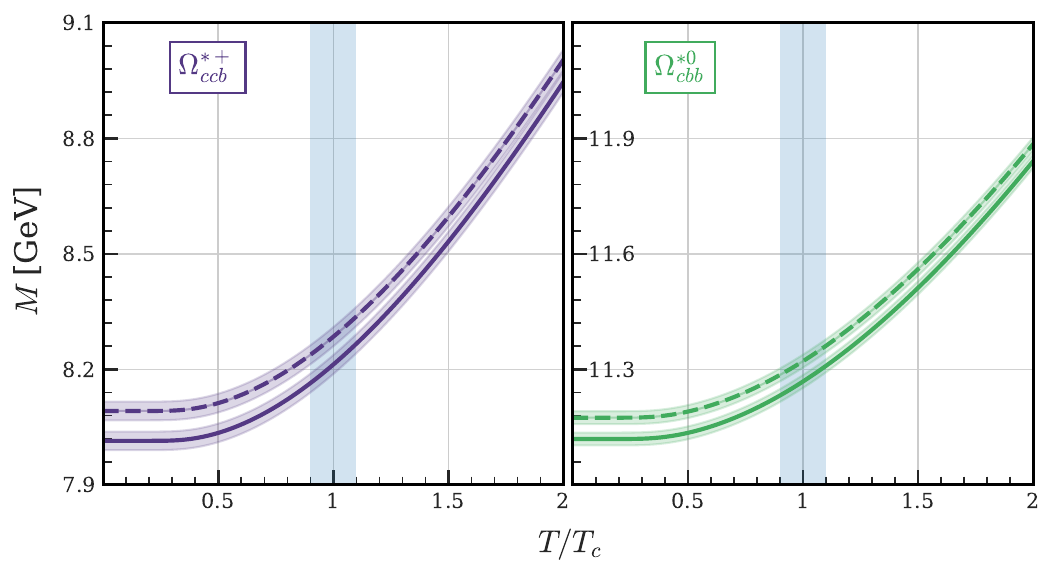}
\vspace{-0.7cm}
    \caption{\justifying 
    Screening masses of the spin-$3/2$ baryons composed exclusively of $\tc$ and $\tb$ quarks. In each case, the positive parity states (solid lines) are compared with their respective parity partners (dashed lines). 
    The results are computed in the temperature range from $T=0$ MeV up to  $T/T_c=2$. The shaded blue band indicates the crossover region. Uncertainty bands of 0.3\% for $\Omega^{*+}_{\tc\tc\tb}$ and 0.15\% for $\Omega^{*0}_{\tc\tb\tb}$ are shown.
    }
    \label{triply-HH}
\end{figure}
 \begin{table}[htbp]
\centering
\caption{\justifying
Screening masses (in GeV) of spin-3/2 fully-heavy baryons within the CI framework for both positive and negative parity states at $T=0$ MeV and $T=2\,T_c$. The rows labeled $\Delta_{T_0}$ and $\Delta_{T_1}$ report the values of $\Delta_T$ in Eq.~(\ref{eq:delta_percentage}). The last two rows list the corresponding values of $T_H$ and $T_{HC}$ in GeV}.
\begin{tabular}{@{\extracolsep{0.0cm}}c|cc|cc}
\hline
&
\multicolumn{2}{c|}{$\Omega_{ccb}^{*+}$}
&
\multicolumn{2}{c}{$\Omega_{cbb}^{*0}$}
\\[0.1cm]
\hline
&
\, $3/2^+$ \, & \, $3/2^-$ \,
&
\, $3/2^+$ \, & \, $3/2^-$ \,
\\[0.1cm]
\hline
\rule{0ex}{2.5ex}
$T=0$ MeV
&
8.01 & 8.09
&
11.12 & 11.17
\\

\rule{0ex}{2.5ex}
$\Delta_{T_0}$
&
\multicolumn{2}{c|}{0.99}
&
\multicolumn{2}{c}{0.45}
\\
\hline

\rule{0ex}{2.5ex}
$T=2T_c$
&
8.95 & 9.00
&
11.84 & 11.89
\\

\rule{0ex}{2.5ex}
$\Delta_{T_1}$
&
\multicolumn{2}{c|}{0.56}
&
\multicolumn{2}{c}{0.42}
\\
\hline

\rule{0ex}{2.5ex}
$T_H$
&
0.07 & 0.07
&
0.07 & 0.07
\\

\rule{0ex}{2.5ex}
$T_{HC}$
&
0.18 & 0.18
&
0.19 &  0.19
\\
\hline
\end{tabular}
\label{mspyn-heavy}
\end{table}
In contrast to the previous cases, the values of $T_{HC}$ show no 
dependence on the parity channel for the fully-heavy baryons considered 
here. The positive and negative parity states yield identical values 
of $T_{HC}$, and therefore satisfy Eq.~(\ref{rel_thc}) as an equality.\\
These results indicate that, for these fully-heavy systems, the onset 
of strong thermal modifications occurs at the same characteristic 
temperature in both parity channels, suggesting that the parity 
dependence of $T_{HC}$ observed in other flavor sectors becomes 
negligible as the heavy-quark content increases. With this discussion, 
we conclude our analysis of the thermal behavior of the spin-$3/2$ 
baryons considered in this work. The main findings are summarized in 
the following section.\\
\section{SUMMARY AND PERSPECTIVE}
\label{Summary}
In this work, we extended previous finite-temperature studies of light spin-$3/2$ baryons within the CI framework to a broad class of systems containing strange, charm, and bottom quarks. Our analysis comprises 16 baryonic
flavour channels
spanning the light, heavy-light, and fully-heavy
sectors, including both positive and negative parity
states. We also investigated the thermal evolution
of the corresponding squared Faddeev amplitudes,
providing insight into the persistence of diquark
correlations and the behavior of parity partners
in a hot medium. The study covers temperatures from $T = 0$ MeV up to 
$T/T_c=2$, with the specific temperature range depending on the 
baryon sector considered. Our main findings can be summarized as follows:
\begin{itemize}
\item The $\Delta^{++}$ baryon provides a representative example of the 
thermal evolution observed in the present study. Its screening masses 
exhibit an overall increase with temperature and, at high temperatures, 
display a behavior consistent with an approach toward the asymptotic free 
theory limit, $3\pi T$ (Fig.~\ref{fig:uuu-p}). In the light sector, where 
comparisons with previous calculations are available, we find good overall 
agreement with the results of Ref.~\cite{Chen:2024emt}. A similar overall 
increase with temperature is observed for the spin-$3/2$ baryons considered 
here, although the detailed behavior near the crossover depends on the 
parity channel and thermal effects become progressively weaker as the 
heavy-quark content of the system increases 
(Figs.~\ref{triply-LL}, \ref{triply-LH}, \ref{udosq}, and 
\ref{triply-HH}).
\item The thermal evolution of the relative percentage difference 
between positive and negative parity states exhibits a clear dependence 
on the flavor composition of the baryon. The quantitative results reported 
in Tables~\ref{tab:screening_25}, \ref{table-us}, \ref{mspyn}, and 
\ref{mspyn-heavy} show that $\Delta_T$ decreases with increasing 
temperature in most of the channels considered, indicating a tendency 
toward a reduced relative parity splitting. However, this behavior is 
not universal, since some baryons exhibit a mild increase of $\Delta_T$ 
over the temperature range considered. In particular, among the heavy 
baryons in Table~\ref{mspyn}, this occurs for the 
$\Omega_c^{*0}$, $\Omega_{cc}^{*+}$, and $\Omega_b^{*-}$ states. 
Nevertheless, the uncertainty bands associated with the screening masses, 
as illustrated in Fig.~\ref{udosq}, show that moderate variations of the 
model parameters can modify the separation between the parity partners 
and allow their masses to approach more closely at high temperatures. 
Therefore, while the central values do not exhibit a universal monotonic 
reduction of the relative parity splitting, the results remain compatible, 
within the model uncertainties, with a stronger tendency toward parity 
degeneracy at high temperature. Overall, these findings indicate that the 
thermal evolution of the parity partners is flavor dependent and sensitive 
to the uncertainties of the model parameters.
\item The squared Faddeev amplitudes generally exhibit a mild temperature 
dependence below the crossover region, although their thermal evolution 
varies across the different baryon channels. At higher temperatures, changes 
in the relative strength of the diquark correlations become more apparent. 
Nevertheless, the results shown in Figs.~\ref{triply-LH}, \ref{amp-LH}, and 
\ref{udosq3} demonstrate that the dominant diquark correlations remain clearly 
identifiable throughout the temperature range considered, supporting the 
persistence of quark-diquark correlations even above the crossover region.
\item To quantify the thermal evolution, we introduced the characteristic 
temperatures $T_H$ and $T_{HC}$ in Eq.~\eqref{theq}, associated with the 
onset of visible and strong thermal modifications, respectively. The results 
collected in Tables~\ref{triply-TH}, \ref{table-us}, \ref{mspyn}, and 
\ref{mspyn-heavy} reveal the general trend
\begin{equation}
T_{HC}^{3/2^-} \gtrsim T_{HC}^{3/2^+},
\end{equation}
indicating that negative parity states generally require comparable or 
slightly higher temperatures than their positive parity counterparts for 
the onset of strong thermal modifications. Among all the baryons considered, 
only two states deviate from this trend: the $\Xi^{*0}$, for which 
$T_{HC}^{3/2^-}$ is lower than $T_{HC}^{3/2^+}$ by $0.05$~GeV, and the 
$\Omega_c^{*0}$, for which the corresponding difference is $0.02$~GeV. 
For the fully-heavy $\Omega_{ccb}^{*+}$ and $\Omega_{cbb}^{*0}$ baryons, 
the positive and negative parity states yield identical values of $T_{HC}$, 
indicating that the parity dependence of this characteristic temperature 
becomes negligible in these systems.
\item Comparison with the critical degeneracy temperatures reported in 
Ref.~\cite{Chen:2024emt} reveals a clear hierarchy in the thermal evolution 
of baryons. Small thermal modifications appear first, followed by stronger 
thermal changes near the crossover region, while the convergence of the 
positive and negative parity states occurs only at significantly higher 
temperatures. This comparison highlights the flavor dependence of the 
different thermal scales, particularly as the strangeness content of the 
baryon increases, while our results further show that the thermal response 
becomes progressively weaker as the heavy-quark content of the baryon 
increases.

\item Contrasting our results with those of Ref.~\cite{Chen:2024emt} reveals that different 
treatments of the Faddeev kernel can produce noticeably different thermal 
evolutions of the Faddeev amplitudes while yielding rather similar screening 
masses for the positive parity baryons, as illustrated in 
Fig.~\ref{amp-LH}. For the negative parity states, however, the differences 
between the two approaches become more pronounced. In both parity sectors, 
these discrepancies are primarily confined to temperatures below the 
corresponding characteristic temperature $T_{HC}$, where strong thermal 
modifications begin to develop. Above this scale, the results obtained with 
the different kernels progressively converge, indicating a reduced 
sensitivity to the details of the Faddeev kernel. These observations suggest 
that screening masses are generally robust against moderate variations in 
the internal structure of the baryon, whereas the Faddeev amplitudes provide 
a more sensitive probe of the underlying dynamics.

\end{itemize}

Overall, the present work extends previous finite- temperature CI studies 
to the spin-$3/2$ baryon sector, providing a unified description of light, 
heavy, and fully-heavy baryons in a hot medium. The results offer new 
insight into the interplay between heavy-quark dynamics, chiral symmetry 
restoration, and the persistence of diquark correlations under extreme 
thermal conditions.\\
A natural extension of the present analysis is a systematic investigation 
of the screening masses of spin-$1/2$ baryons at finite- temperature. Such 
a study would allow a direct comparison between the thermal evolution of 
the octet and decuplet sectors within the same symmetry-preserving 
framework. However, the corresponding Faddeev equation involves a 
substantially more complex coupled system, leading to an $8\times8$ kernel 
matrix whose treatment is considerably more demanding from both 
computational and theoretical perspectives. We therefore defer this 
investigation to future work.

\begin{acknowledgements}
\vspace*{-2mm}
The authors thank the Referee for the careful reading of the
manuscript and for the valuable comments and suggestions, which helped
to improve the clarity and overall presentation of this work.
The authors also thank M.~A.~Ramírez-Garrido
for valuable assistance with the calculations presented in this work.
L.~X.~Guti\'errez-Guerrero acknowledges the {\em Secretaría de Ciencia,
Humanidades, Tecnología e Innovación} (SECIHTI) for the support provided
through the {\em Investigadoras e Investigadores por México} program and
Project CBF2023-2024-268, ``Hadronic Physics at JLab: Deciphering the
Internal Structure of Mesons and Baryons,'' funded under the 2023--2024
Frontier Science Call. The work of R.~J.~Hern\'andez-Pinto is partly
supported by SECIHTI (Mexico) through the {\em Sistema Nacional de
Investigadoras e Investigadores}.
\end{acknowledgements}


\appendix
\setcounter{equation}{0}
\renewcommand{\theequation}{\Alph{section}.\arabic{equation}}
\section{Euclidean Conventions}
\label{App:EM}
In our Euclidean formulation:
\begin{equation}
p\cdot q=\sum_{i=1}^4 p_i q_i\,,
\end{equation}
and
\begin{eqnarray}\nn
&&\{\gamma_\mu,\gamma_\nu\}=2\,\delta_{\mu\nu}\,;\;
\gamma_\mu^\dagger = \gamma_\mu\,;\;
\sigma_{\mu\nu}= \frac{i}{2}[\gamma_\mu,\gamma_\nu]\,; \; \\
&&{\rm tr}\,[\gamma_5\gamma_\mu\gamma_\nu\gamma_\rho\gamma_\sigma]=
-4\,\epsilon_{\mu\nu\rho\sigma}\,, \epsilon_{1234}= 1\,.
\end{eqnarray}
A positive energy spinor satisfies
\begin{equation}
\bar u(P,r)\, (i \gamma\cdot P + M) = 0 = (i\gamma\cdot P + M)\, u(P,r)\,,
\end{equation}
where $r=\pm$ is the spin label.  It is normalized as~:
\begin{equation}
\bar u(P,r) \, u(P,r) = 2 M \,,
\end{equation}
and may be expressed explicitly as~:
\begin{equation}
u(P,r) = \sqrt{M- i {\cal E}}
\left(
\begin{array}{c}
\chi_r\\
\displaystyle \frac{\vec{\sigma}\cdot \vec{P}}{M - i {\cal E}} \chi_r
\end{array}
\right)\,,
\end{equation}
with ${\cal E} = i \sqrt{\vec{P}^2 + M^2}$, and
\begin{equation}
\chi_+ = \left( \begin{array}{c} 1 \\ 0  \end{array}\right)\,,\;
\chi_- = \left( \begin{array}{c} 0\\ 1  \end{array}\right)\,.
\end{equation}
For the free-particle spinor, $\bar u(P,r)= u(P,r)^\dagger \gamma_4$.
It can be used to construct a positive energy projection operator:
\begin{equation}
\label{Lplus} \Lambda_+(P)\equiv \frac{1}{2 M}\,\sum_{r=\pm} \, u(P,r) \, \bar
u(P,r) = \frac{1}{2M} \left( -i \gamma\cdot P + M\right).
\end{equation}
On the other hand, a negative energy spinor satisfies
\begin{equation}
\bar v(P,r)\,(i\gamma\cdot P - M) = 0 = (i\gamma\cdot P - M) \, v(P,r)\,,
\end{equation}
and possesses properties and constraints obtained through obvious analogy
with $u(P,r)$. A charge-conjugated BSA is obtained via
\begin{equation}
\label{chargec}
\bar\Gamma(k;P) = C^\dagger \, \Gamma(-k;P)^{\mathbb T}\,C\,;
\end{equation}
where, $\mathbb{T}$ denotes transposing  all matrix indices and
$C=\gamma_2\gamma_4$ is the charge conjugation matrix, $C^\dagger=-C$.  Moreover, we note that
\begin{equation}
C^\dagger \gamma_\mu^{\mathbb{T}} \, C = - \gamma_\mu\,, \; [C,\gamma_5] = 0\,.
\end{equation}
We employ a Rarita-Schwinger spinor to represent a covariant spin-$3/2$ field.  The positive energy
spinor is defined by the following equations:
\begin{equation}
\label{rarita}
(i \gamma\cdot P + M)\, u_\mu(P,s) = 0\,,
\gamma_\mu u_\mu(P,s) = 0\,,
P_\mu u_\mu(P,s) = 0,
\end{equation}
where $s=\{\pm3/2,\pm1/2\}$.  It is normalized as:
\begin{equation}
\bar u_{\mu}(P,s^\prime) \, u_\mu(P,s) = 2 M \delta_{s's} \,,
\end{equation}
and satisfies a completeness relation
\begin{equation}
\label{Deltacomplete}
\frac{1}{2 M}\sum_{s=-3/2}^{+3/2} u_\mu(P,s)\,\bar{u}_\nu(P,s) =
\Lambda_+(P)\,R_{\mu\nu}\,,
\end{equation}
where
\begin{equation}
R_{\mu\nu} = \delta_{\mu\nu} \mathbb{I}_{\rm D} -\frac{1}{3} \gamma_\mu \gamma_\nu +
\frac{2}{3} \hat P_\mu \hat P_\nu \mathbb{I}_{\rm D} - i\frac{1}{3} [ \hat P_\mu
\gamma_\nu - \hat P_\nu \gamma_\mu]\,,
\label{Rde}\end{equation}
with $\hat P^2 = -1$. It is very useful in simplifying the FE for a positive energy decouplet state.
\section{Flavor diquarks}
\label{App:Flavdiq}
We define the following set of flavor column matrices,
\begin{equation}\nn
t^{\tu}=\begin{pmatrix} 1  \\ 0  \\ 0 \\ 0 \\0 \end{pmatrix},\;\;\;\;
t^{\td}=\begin{pmatrix} 0  \\ 1  \\ 0  \\ 0\\0  \end{pmatrix},\;\;\;\;
t^{\ts}=\begin{pmatrix} 0  \\ 0  \\ 1 \\ 0 \\0  \end{pmatrix},\;\;\;\;
\end{equation}
\begin{equation}
t^{\tc}=\begin{pmatrix} 0  \\ 0  \\ 0 \\ 1\\0  \end{pmatrix},\;\;\;\;
t^{\tb}=\begin{pmatrix} 0  \\ 0  \\ 0 \\ 0\\1  \end{pmatrix},\;\;\;\;
\end{equation}
and
\bea \nn
t^f=\left[\begin{matrix} \nn
t^{\tq T}t^{\{\tqu \tq\}}t^{\{\tqu\tq\}T}t^{\tq} &&t^{\tq T}t^{\{\tq\tq\}}t^{\{\tqu\tq\}T}t^{\tqu}\\ \\
t^{\tqu T}t^{\{\tqu\tq\}}t^{\{\tq\tq\}T}t^{\tq} && t^{\tqu T}t^{\{\tq\tq\}}t^{\{\tq\tq\}T}t^{\tqu}
\end{matrix}\right] \,.
\eea
The flavor matrices for the diquarks are
\begin{equation}\nonumber
\begin{array}{cc}
\bf{t^{\{\tu\tu\}}=\begin{pmatrix} \sqrt{2} & 0 & 0& 0 &0\\ 0 & 0 & 0& 0&0 \\ 0 &  0 & 0 & 0&0 \\ 0 &  0 & 0 & 0&0\\ 0 & 0 & 0& 0& 0  \end{pmatrix}}\,,
&
\bf{t^{\{\tu\td\}}=\begin{pmatrix} 0 & 1 & 0 & 0&0\\ 1 & 0 & 0& 0&0 \\ 0 &  0 & 0& 0&0 \\ 0 &  0 & 0 & 0 &0\\ 0 & 0 & 0& 0& 0  \end{pmatrix}},\\[10ex]
\bf{t^{\{\tu\ts\}}=\begin{pmatrix} 0 & 0 & 1& 0&0  \\ 0 & 0 & 0& 0&0  \\ 1 &  0 & 0& 0 &0 \\ 0 &  0 & 0 & 0&0 \\ 0 & 0 & 0& 0& 0   \end{pmatrix}}\,,
&
\bf{t^{\{\tu\tc\}}}=\begin{pmatrix} 0 & 0 & 0& 1&0 \\ 0 & 0 & 0 & 0&0\\ 0 &  0 & 0 & 0&0\\  1 &  0 & 0 & 0&0\\ 0 & 0 & 0& 0& 0  \end{pmatrix},
\\[10ex]
%
\bf{t^{\{\td\td\}}=\begin{pmatrix} 0 & 0 & 0 & 0 & 0 \\ 0 &  \sqrt{2} & 0& 0 & 0 \\ 0 &  0 & 0& 0& 0  \\ 0 & 0 & 0 & 0& 0\\ 0 & 0 & 0& 0& 0  \end{pmatrix}}\,,
&
\bf{t^{\{\td\ts\}}}=\begin{pmatrix} 0 & 0 & 0& 0 & 0 \\ 0 & 0 & 1 & 0 & 0\\ 0 &  1 & 0 & 0 & 0\\ 0 & 0 & 0 & 0& 0\\ 0 & 0 & 0& 0& 0   \end{pmatrix},
\\ [10ex]
\bf{t^{\{\ts\ts\}}=\begin{pmatrix} 0 & 0 & 0& 0& 0 \\ 0 & 0 & 0& 0& 0 \\ 0 &  0 & \sqrt{2} & 0 & 0 \\ 0 & 0 & 0 & 0 & 0\\ 0 & 0 & 0& 0& 0  \end{pmatrix}}, 
&
\bf{t^{\{\tc\tc\}}}=\begin{pmatrix} 0 & 0 & 0& 0 & 0 \\ 0 & 0 & 0 & 0 & 0\\ 0 &  0 & 0 & 0 & 0\\  0 &  0 & 0 &  \sqrt{2} & 0\\ 0 & 0 & 0& 0& 0   \end{pmatrix}\,,
\\[10ex]
\bf{t^{\{\td\tc\}}}=\begin{pmatrix} 0 & 0 & 0& 0 & 0\\ 0 & 0 & 0 & 1& 0\\ 0 &  0 & 0 & 0& 0\\  0 &  1& 0 & 0& 0 \\ 0 & 0 & 0& 0& 0   \end{pmatrix},
&
\bf{t^{\{\ts\tc\}}}=\begin{pmatrix} 0 & 0 & 0& 0 & 0 \\ 0 & 0 & 0 & 0 & 0\\ 0 &  0 & 0 & 1 & 0\\  0 &  0& 1 & 0 & 0 \\ 0 & 0 & 0& 0& 0   \end{pmatrix},\\
[10ex]
\bf{t^{\{\tb\tb\}}}=\begin{pmatrix} 0 & 0 & 0& 0& 0\\ 0 & 0 & 0 & 0& 0\\ 0 &  0 & 0 & 0& 0\\  0 &  0& 0 & 0& 0 \\ 0 & 0 & 0& 0&  \sqrt{2}    \end{pmatrix}\,,
&
\bf{t^{\{\tb\tu\}}}=\begin{pmatrix} 0 & 0 & 0& 0& 1\\ 0 & 0 & 0 & 0& 0\\ 0 &  0 & 0 & 0& 0\\  0 &  0& 0 & 0& 0 \\ 1 & 0 & 0& 0& 0 \end{pmatrix}, \\
 [10ex]
 \end{array}
\end{equation}
\begin{equation}\nonumber
\begin{array}{cc}
\bf{t^{\{\tb\td\}}}=\begin{pmatrix} 0 & 0 & 0& 0& 0\\ 0 & 0 & 0 & 0& 1\\ 0 &  0 & 0 & 0& 0\\  0 &  0& 0 & 0& 0 \\ 0 & 1 & 0& 0& 0   \end{pmatrix} \,,
&
\bf{t^{\{\tb\tc\}}}=\begin{pmatrix} 0 & 0 & 0& 0& 0\\ 0 & 0 & 0 & 0& 0\\ 0 &  0 & 0 & 0& 0\\  0 &  0& 0 & 0& 1 \\ 0 & 0 & 0& 1& 0   \end{pmatrix},
\\[10ex]
\bf{t^{\{\tb\ts\}}}=\begin{pmatrix} 0 & 0 & 0& 0& 0\\ 0 & 0 & 0 & 0& 0\\ 0 &  0 & 0 & 0& 1\\  0 &  0& 0 & 0& 0 \\ 0 & 0 & 1& 0& 0   
 \end{pmatrix}\,.
 \end{array}
\end{equation}
\bibliography{ccc-a}

\begin{thebibliography}{50}
\expandafter\ifx\csname natexlab\endcsname\relax\def\natexlab#1{#1}\fi
\expandafter\ifx\csname bibnamefont\endcsname\relax
  \def\bibnamefont#1{#1}\fi
\expandafter\ifx\csname bibfnamefont\endcsname\relax
  \def\bibfnamefont#1{#1}\fi
\expandafter\ifx\csname citenamefont\endcsname\relax
  \def\citenamefont#1{#1}\fi
\expandafter\ifx\csname url\endcsname\relax
  \def\url#1{\texttt{#1}}\fi
\expandafter\ifx\csname urlprefix\endcsname\relax\def\urlprefix{URL }\fi
\providecommand{\bibinfo}[2]{#2}
\providecommand{\eprint}[2][]{\url{#2}}

\bibitem[{\citenamefont{Messchendorp et~al.}(2025)}]{Messchendorp:2025men}
\bibinfo{author}{\bibfnamefont{J.~G.} \bibnamefont{Messchendorp}}
  \bibnamefont{et~al.} (\bibinfo{year}{2025}), \eprint{2512.15986}.

\bibitem[{\citenamefont{Cheng et~al.}(2011)}]{Cheng:2010fe}
\bibinfo{author}{\bibfnamefont{M.}~\bibnamefont{Cheng}} \bibnamefont{et~al.},
  \bibinfo{journal}{Eur. Phys. J. C} \textbf{\bibinfo{volume}{71}},
  \bibinfo{pages}{1564} (\bibinfo{year}{2011}), \eprint{1010.1216}.

\bibitem[{\citenamefont{Mukherjee}(2009)}]{Mukherjee:2008tr}
\bibinfo{author}{\bibfnamefont{S.}~\bibnamefont{Mukherjee}},
  \bibinfo{journal}{Nucl. Phys. A} \textbf{\bibinfo{volume}{820}},
  \bibinfo{pages}{283C} (\bibinfo{year}{2009}), \eprint{0810.2906}.

\bibitem[{\citenamefont{Bazavov and Weber}(2021)}]{Bazavov:2020teh}
\bibinfo{author}{\bibfnamefont{A.}~\bibnamefont{Bazavov}} \bibnamefont{and}
  \bibinfo{author}{\bibfnamefont{J.~H.} \bibnamefont{Weber}},
  \bibinfo{journal}{Prog. Part. Nucl. Phys.} \textbf{\bibinfo{volume}{116}},
  \bibinfo{pages}{103823} (\bibinfo{year}{2021}), \eprint{2010.01873}.

\bibitem[{\citenamefont{Petreczky}(2009)}]{Petreczky:2009at}
\bibinfo{author}{\bibfnamefont{P.}~\bibnamefont{Petreczky}},
  \bibinfo{journal}{Nucl. Phys. A} \textbf{\bibinfo{volume}{830}},
  \bibinfo{pages}{11C} (\bibinfo{year}{2009}), \eprint{0908.1917}.

\bibitem[{\citenamefont{Song}(1994)}]{Song:1993ipa}
\bibinfo{author}{\bibfnamefont{C.}~\bibnamefont{Song}}, \bibinfo{journal}{Phys.
  Rev. D} \textbf{\bibinfo{volume}{49}}, \bibinfo{pages}{1556}
  (\bibinfo{year}{1994}).

\bibitem[{\citenamefont{Chen et~al.}(2025)\citenamefont{Chen, Gao, and
  Qin}}]{Chen:2024emt}
\bibinfo{author}{\bibfnamefont{C.}~\bibnamefont{Chen}},
  \bibinfo{author}{\bibfnamefont{F.}~\bibnamefont{Gao}}, \bibnamefont{and}
  \bibinfo{author}{\bibfnamefont{S.-x.} \bibnamefont{Qin}},
  \bibinfo{journal}{Phys. Rev. D} \textbf{\bibinfo{volume}{112}},
  \bibinfo{pages}{014022} (\bibinfo{year}{2025}), \eprint{2412.15045}.

\bibitem[{\citenamefont{Wang et~al.}(2013)\citenamefont{Wang, Liu, Chang,
  Roberts, and Schmidt}}]{Wang:2013wk}
\bibinfo{author}{\bibfnamefont{K.-l.} \bibnamefont{Wang}},
  \bibinfo{author}{\bibfnamefont{Y.-x.} \bibnamefont{Liu}},
  \bibinfo{author}{\bibfnamefont{L.}~\bibnamefont{Chang}},
  \bibinfo{author}{\bibfnamefont{C.~D.} \bibnamefont{Roberts}},
  \bibnamefont{and} \bibinfo{author}{\bibfnamefont{S.~M.}
  \bibnamefont{Schmidt}}, \bibinfo{journal}{Phys. Rev. D}
  \textbf{\bibinfo{volume}{87}}, \bibinfo{pages}{074038}
  (\bibinfo{year}{2013}), \eprint{1301.6762}.

\bibitem[{\citenamefont{Ram{\'\i}rez-Garrido
  et~al.}(2025)\citenamefont{Ram{\'\i}rez-Garrido, Hern{\'a}ndez-Pinto,
  Higuera-Angulo, and Guti{\'e}rrez-Guerrero}}]{Ramirez-Garrido:2025rsu}
\bibinfo{author}{\bibfnamefont{M.~A.} \bibnamefont{Ram{\'\i}rez-Garrido}},
  \bibinfo{author}{\bibfnamefont{R.~J.} \bibnamefont{Hern{\'a}ndez-Pinto}},
  \bibinfo{author}{\bibfnamefont{I.~M.} \bibnamefont{Higuera-Angulo}},
  \bibnamefont{and} \bibinfo{author}{\bibfnamefont{L.~X.}
  \bibnamefont{Guti{\'e}rrez-Guerrero}}, \bibinfo{journal}{Phys. Rev. D}
  \textbf{\bibinfo{volume}{112}}, \bibinfo{pages}{034024}
  (\bibinfo{year}{2025}), \eprint{2508.01099}.

\bibitem[{\citenamefont{Guti{\'e}rrez-Guerrero
  et~al.}(2026)\citenamefont{Guti{\'e}rrez-Guerrero, Ram{\'\i}rez-Garrido,
  de~Le{\'o}n, and Hern{\'a}ndez-Pinto}}]{Gutierrez-Guerrero:2026hhu}
\bibinfo{author}{\bibfnamefont{L.~X.} \bibnamefont{Guti{\'e}rrez-Guerrero}},
  \bibinfo{author}{\bibfnamefont{M.~A.} \bibnamefont{Ram{\'\i}rez-Garrido}},
  \bibinfo{author}{\bibfnamefont{M.~A.~P.} \bibnamefont{de~Le{\'o}n}},
  \bibnamefont{and} \bibinfo{author}{\bibfnamefont{R.~J.}
  \bibnamefont{Hern{\'a}ndez-Pinto}}, \bibinfo{journal}{Phys. Rev. D}
  \textbf{\bibinfo{volume}{113}}, \bibinfo{pages}{014035}
  (\bibinfo{year}{2026}), \eprint{2601.04332}.

\bibitem[{\citenamefont{Sumit et~al.}(2023)\citenamefont{Sumit, Haque, and
  Patra}}]{Sumit:2023hjj}
\bibinfo{author}{\bibnamefont{Sumit}},
  \bibinfo{author}{\bibfnamefont{N.}~\bibnamefont{Haque}}, \bibnamefont{and}
  \bibinfo{author}{\bibfnamefont{B.~K.} \bibnamefont{Patra}},
  \bibinfo{journal}{Phys. Lett. B} \textbf{\bibinfo{volume}{845}},
  \bibinfo{pages}{138143} (\bibinfo{year}{2023}), \eprint{2305.08525}.

\bibitem[{\citenamefont{Bazavov et~al.}(2019)}]{Bazavov:2019www}
\bibinfo{author}{\bibfnamefont{A.}~\bibnamefont{Bazavov}} \bibnamefont{et~al.},
  \bibinfo{journal}{Phys. Rev. D} \textbf{\bibinfo{volume}{100}},
  \bibinfo{pages}{094510} (\bibinfo{year}{2019}), \eprint{1908.09552}.

\bibitem[{\citenamefont{Hashimoto et~al.}(1993)\citenamefont{Hashimoto,
  Nakamura, and Stamatescu}}]{Hashimoto:1992np}
\bibinfo{author}{\bibfnamefont{T.}~\bibnamefont{Hashimoto}},
  \bibinfo{author}{\bibfnamefont{A.}~\bibnamefont{Nakamura}}, \bibnamefont{and}
  \bibinfo{author}{\bibfnamefont{I.~O.} \bibnamefont{Stamatescu}},
  \bibinfo{journal}{Nucl. Phys. B} \textbf{\bibinfo{volume}{400}},
  \bibinfo{pages}{267} (\bibinfo{year}{1993}).

\bibitem[{\citenamefont{Beraudo et~al.}(2018)}]{Rapp:2018qla}
\bibinfo{author}{\bibfnamefont{A.}~\bibnamefont{Beraudo}} \bibnamefont{et~al.},
  \bibinfo{journal}{Nucl. Phys. A} \textbf{\bibinfo{volume}{979}},
  \bibinfo{pages}{21} (\bibinfo{year}{2018}), \eprint{1803.03824}.

\bibitem[{\citenamefont{Dong et~al.}(2019)\citenamefont{Dong, Lee, and
  Rapp}}]{Dong:2019byy}
\bibinfo{author}{\bibfnamefont{X.}~\bibnamefont{Dong}},
  \bibinfo{author}{\bibfnamefont{Y.-J.} \bibnamefont{Lee}}, \bibnamefont{and}
  \bibinfo{author}{\bibfnamefont{R.}~\bibnamefont{Rapp}},
  \bibinfo{journal}{Ann. Rev. Nucl. Part. Sci.} \textbf{\bibinfo{volume}{69}},
  \bibinfo{pages}{417} (\bibinfo{year}{2019}), \eprint{1903.07709}.

\bibitem[{\citenamefont{Adamov{\'a} et~al.}(2019)}]{Adamova:2019vkf}
\bibinfo{author}{\bibfnamefont{D.}~\bibnamefont{Adamov{\'a}}}
  \bibnamefont{et~al.} (\bibinfo{year}{2019}), \eprint{1902.01211}.

\bibitem[{\citenamefont{Li et~al.}(2025{\natexlab{a}})\citenamefont{Li, Liu,
  Zheng, and Chen}}]{Li:2025nca}
\bibinfo{author}{\bibfnamefont{T.}~\bibnamefont{Li}},
  \bibinfo{author}{\bibfnamefont{J.}~\bibnamefont{Liu}},
  \bibinfo{author}{\bibfnamefont{S.}~\bibnamefont{Zheng}}, \bibnamefont{and}
  \bibinfo{author}{\bibfnamefont{B.}~\bibnamefont{Chen}},
  \bibinfo{journal}{Phys. Lett. B} \textbf{\bibinfo{volume}{868}},
  \bibinfo{pages}{139738} (\bibinfo{year}{2025}{\natexlab{a}}),
  \eprint{2503.11223}.

\bibitem[{\citenamefont{Linde}(1980)}]{Linde:1980ts}
\bibinfo{author}{\bibfnamefont{A.~D.} \bibnamefont{Linde}},
  \bibinfo{journal}{Phys. Lett. B} \textbf{\bibinfo{volume}{96}},
  \bibinfo{pages}{289} (\bibinfo{year}{1980}).

\bibitem[{\citenamefont{Bazavov et~al.}(2012)}]{Bazavov:2011nk}
\bibinfo{author}{\bibfnamefont{A.}~\bibnamefont{Bazavov}} \bibnamefont{et~al.},
  \bibinfo{journal}{Phys. Rev. D} \textbf{\bibinfo{volume}{85}},
  \bibinfo{pages}{054503} (\bibinfo{year}{2012}), \eprint{1111.1710}.

\bibitem[{\citenamefont{Giusti et~al.}(2024)\citenamefont{Giusti, Harris,
  Laudicina, Pepe, and Rescigno}}]{Giusti:2024ohu}
\bibinfo{author}{\bibfnamefont{L.}~\bibnamefont{Giusti}},
  \bibinfo{author}{\bibfnamefont{T.}~\bibnamefont{Harris}},
  \bibinfo{author}{\bibfnamefont{D.}~\bibnamefont{Laudicina}},
  \bibinfo{author}{\bibfnamefont{M.}~\bibnamefont{Pepe}}, \bibnamefont{and}
  \bibinfo{author}{\bibfnamefont{P.}~\bibnamefont{Rescigno}},
  \bibinfo{journal}{Phys. Lett. B} \textbf{\bibinfo{volume}{855}},
  \bibinfo{pages}{138799} (\bibinfo{year}{2024}), \eprint{2405.04182}.

\bibitem[{\citenamefont{Detar and Kogut}(1987)}]{Detar:1987hib}
\bibinfo{author}{\bibfnamefont{C.~E.} \bibnamefont{Detar}} \bibnamefont{and}
  \bibinfo{author}{\bibfnamefont{J.~B.} \bibnamefont{Kogut}},
  \bibinfo{journal}{Phys. Rev. D} \textbf{\bibinfo{volume}{36}},
  \bibinfo{pages}{2828} (\bibinfo{year}{1987}).

\bibitem[{\citenamefont{Gocksch et~al.}(1988)\citenamefont{Gocksch, Rossi, and
  Heller}}]{Gocksch:1987nt}
\bibinfo{author}{\bibfnamefont{A.}~\bibnamefont{Gocksch}},
  \bibinfo{author}{\bibfnamefont{P.}~\bibnamefont{Rossi}}, \bibnamefont{and}
  \bibinfo{author}{\bibfnamefont{U.~M.} \bibnamefont{Heller}},
  \bibinfo{journal}{Phys. Lett. B} \textbf{\bibinfo{volume}{205}},
  \bibinfo{pages}{334} (\bibinfo{year}{1988}).

\bibitem[{\citenamefont{Eletsky and Ioffe}(1988)}]{Eletsky:1988an}
\bibinfo{author}{\bibfnamefont{V.~L.} \bibnamefont{Eletsky}} \bibnamefont{and}
  \bibinfo{author}{\bibfnamefont{B.~L.} \bibnamefont{Ioffe}},
  \bibinfo{journal}{Sov. J. Nucl. Phys.} \textbf{\bibinfo{volume}{48}},
  \bibinfo{pages}{384} (\bibinfo{year}{1988}).

\bibitem[{\citenamefont{Azizi and Bozk{\i}r}(2016)}]{Azizi:2016ddw}
\bibinfo{author}{\bibfnamefont{K.}~\bibnamefont{Azizi}} \bibnamefont{and}
  \bibinfo{author}{\bibfnamefont{G.}~\bibnamefont{Bozk{\i}r}},
  \bibinfo{journal}{Eur. Phys. J. C} \textbf{\bibinfo{volume}{76}},
  \bibinfo{pages}{521} (\bibinfo{year}{2016}), \eprint{1606.05452}.

\bibitem[{\citenamefont{Xu et~al.}(2015)\citenamefont{Xu, Liu, and
  Huang}}]{Xu:2015jxa}
\bibinfo{author}{\bibfnamefont{Y.-J.} \bibnamefont{Xu}},
  \bibinfo{author}{\bibfnamefont{Y.-L.} \bibnamefont{Liu}}, \bibnamefont{and}
  \bibinfo{author}{\bibfnamefont{M.-Q.} \bibnamefont{Huang}},
  \bibinfo{journal}{Commun. Theor. Phys.} \textbf{\bibinfo{volume}{63}},
  \bibinfo{pages}{209} (\bibinfo{year}{2015}), \eprint{1502.00429}.

\bibitem[{\citenamefont{Aarts et~al.}(2019)\citenamefont{Aarts, Allton,
  De~Boni, and J{\"a}ger}}]{Aarts:2018glk}
\bibinfo{author}{\bibfnamefont{G.}~\bibnamefont{Aarts}},
  \bibinfo{author}{\bibfnamefont{C.}~\bibnamefont{Allton}},
  \bibinfo{author}{\bibfnamefont{D.}~\bibnamefont{De~Boni}}, \bibnamefont{and}
  \bibinfo{author}{\bibfnamefont{B.}~\bibnamefont{J{\"a}ger}},
  \bibinfo{journal}{Phys. Rev. D} \textbf{\bibinfo{volume}{99}},
  \bibinfo{pages}{074503} (\bibinfo{year}{2019}), \eprint{1812.07393}.

\bibitem[{\citenamefont{Gutierrez-Guerrero
  et~al.}(2010)\citenamefont{Gutierrez-Guerrero, Bashir, Cloet, and
  Roberts}}]{GutierrezGuerrero:2010md}
\bibinfo{author}{\bibfnamefont{L.~X.} \bibnamefont{Gutierrez-Guerrero}},
  \bibinfo{author}{\bibfnamefont{A.}~\bibnamefont{Bashir}},
  \bibinfo{author}{\bibfnamefont{I.~C.} \bibnamefont{Cloet}}, \bibnamefont{and}
  \bibinfo{author}{\bibfnamefont{C.~D.} \bibnamefont{Roberts}},
  \bibinfo{journal}{Phys. Rev.} \textbf{\bibinfo{volume}{C81}},
  \bibinfo{pages}{065202} (\bibinfo{year}{2010}), \eprint{1002.1968}.

\bibitem[{\citenamefont{Aguilar et~al.}(2018)\citenamefont{Aguilar, Binosi,
  Figueiredo, and Papavassiliou}}]{Aguilar:2017dco}
\bibinfo{author}{\bibfnamefont{A.~C.} \bibnamefont{Aguilar}},
  \bibinfo{author}{\bibfnamefont{D.}~\bibnamefont{Binosi}},
  \bibinfo{author}{\bibfnamefont{C.~T.} \bibnamefont{Figueiredo}},
  \bibnamefont{and}
  \bibinfo{author}{\bibfnamefont{J.}~\bibnamefont{Papavassiliou}},
  \bibinfo{journal}{Eur. Phys. J.} \textbf{\bibinfo{volume}{C78}},
  \bibinfo{pages}{181} (\bibinfo{year}{2018}), \eprint{1712.06926}.

\bibitem[{\citenamefont{Binosi and Papavassiliou}(2018)}]{Binosi:2017rwj}
\bibinfo{author}{\bibfnamefont{D.}~\bibnamefont{Binosi}} \bibnamefont{and}
  \bibinfo{author}{\bibfnamefont{J.}~\bibnamefont{Papavassiliou}},
  \bibinfo{journal}{Phys. Rev.} \textbf{\bibinfo{volume}{D97}},
  \bibinfo{pages}{054029} (\bibinfo{year}{2018}), \eprint{1709.09964}.

\bibitem[{\citenamefont{Gao et~al.}(2018)\citenamefont{Gao, Qin, Roberts, and
  Rodriguez-Quintero}}]{Gao:2017uox}
\bibinfo{author}{\bibfnamefont{F.}~\bibnamefont{Gao}},
  \bibinfo{author}{\bibfnamefont{S.-X.} \bibnamefont{Qin}},
  \bibinfo{author}{\bibfnamefont{C.~D.} \bibnamefont{Roberts}},
  \bibnamefont{and}
  \bibinfo{author}{\bibfnamefont{J.}~\bibnamefont{Rodriguez-Quintero}},
  \bibinfo{journal}{Phys. Rev.} \textbf{\bibinfo{volume}{D97}},
  \bibinfo{pages}{034010} (\bibinfo{year}{2018}), \eprint{1706.04681}.

\bibitem[{\citenamefont{Guti\'errez-Guerrero
  et~al.}(2021)\citenamefont{Guti\'errez-Guerrero, Paredes-Torres, and
  Bashir}}]{Gutierrez-Guerrero:2021rsx}
\bibinfo{author}{\bibfnamefont{L.~X.} \bibnamefont{Guti\'errez-Guerrero}},
  \bibinfo{author}{\bibfnamefont{G.}~\bibnamefont{Paredes-Torres}},
  \bibnamefont{and} \bibinfo{author}{\bibfnamefont{A.}~\bibnamefont{Bashir}},
  \bibinfo{journal}{Phys. Rev. D} \textbf{\bibinfo{volume}{104}},
  \bibinfo{pages}{094013} (\bibinfo{year}{2021}), \eprint{2109.09058}.

\bibitem[{\citenamefont{Bedolla et~al.}(2015)\citenamefont{Bedolla,
  Cobos-Mart\'inez, and Bashir}}]{Bedolla:2015mpa}
\bibinfo{author}{\bibfnamefont{M.~A.} \bibnamefont{Bedolla}},
  \bibinfo{author}{\bibfnamefont{J.~J.} \bibnamefont{Cobos-Mart\'inez}},
  \bibnamefont{and} \bibinfo{author}{\bibfnamefont{A.}~\bibnamefont{Bashir}},
  \bibinfo{journal}{Phys. Rev.} \textbf{\bibinfo{volume}{D92}},
  \bibinfo{pages}{054031} (\bibinfo{year}{2015}), \eprint{1601.05639}.

\bibitem[{\citenamefont{Bedolla et~al.}(2016)\citenamefont{Bedolla, Raya,
  Cobos-Mart\'inez, and Bashir}}]{Bedolla:2016yxq}
\bibinfo{author}{\bibfnamefont{M.~A.} \bibnamefont{Bedolla}},
  \bibinfo{author}{\bibfnamefont{K.}~\bibnamefont{Raya}},
  \bibinfo{author}{\bibfnamefont{J.~J.} \bibnamefont{Cobos-Mart\'inez}},
  \bibnamefont{and} \bibinfo{author}{\bibfnamefont{A.}~\bibnamefont{Bashir}},
  \bibinfo{journal}{Phys. Rev.} \textbf{\bibinfo{volume}{D93}},
  \bibinfo{pages}{094025} (\bibinfo{year}{2016}), \eprint{1606.03760}.

\bibitem[{\citenamefont{Raya et~al.}(2018)\citenamefont{Raya, Bedolla,
  Cobos-Mart\'\i{}nez, and Bashir}}]{Raya:2017ggu}
\bibinfo{author}{\bibfnamefont{K.}~\bibnamefont{Raya}},
  \bibinfo{author}{\bibfnamefont{M.~A.} \bibnamefont{Bedolla}},
  \bibinfo{author}{\bibfnamefont{J.~J.} \bibnamefont{Cobos-Mart\'\i{}nez}},
  \bibnamefont{and} \bibinfo{author}{\bibfnamefont{A.}~\bibnamefont{Bashir}},
  \bibinfo{journal}{Few Body Syst.} \textbf{\bibinfo{volume}{59}},
  \bibinfo{pages}{133} (\bibinfo{year}{2018}), \eprint{1711.00383}.

\bibitem[{\citenamefont{Gutiérrez-Guerrero
  et~al.}(2019)\citenamefont{Gutiérrez-Guerrero, Bashir, Bedolla, and
  Santopinto}}]{Gutierrez-Guerrero:2019uwa}
\bibinfo{author}{\bibfnamefont{L.~X.} \bibnamefont{Gutiérrez-Guerrero}},
  \bibinfo{author}{\bibfnamefont{A.}~\bibnamefont{Bashir}},
  \bibinfo{author}{\bibfnamefont{M.~A.} \bibnamefont{Bedolla}},
  \bibnamefont{and}
  \bibinfo{author}{\bibfnamefont{E.}~\bibnamefont{Santopinto}},
  \bibinfo{journal}{Phys. Rev.} \textbf{\bibinfo{volume}{D100}},
  \bibinfo{pages}{114032} (\bibinfo{year}{2019}), \eprint{1911.09213}.

\bibitem[{\citenamefont{Yin et~al.}(2019)\citenamefont{Yin, Chen, Krein,
  Roberts, Segovia, and Xu}}]{Yin:2019bxe}
\bibinfo{author}{\bibfnamefont{P.-L.} \bibnamefont{Yin}},
  \bibinfo{author}{\bibfnamefont{C.}~\bibnamefont{Chen}},
  \bibinfo{author}{\bibfnamefont{G.}~\bibnamefont{Krein}},
  \bibinfo{author}{\bibfnamefont{C.~D.} \bibnamefont{Roberts}},
  \bibinfo{author}{\bibfnamefont{J.}~\bibnamefont{Segovia}}, \bibnamefont{and}
  \bibinfo{author}{\bibfnamefont{S.-S.} \bibnamefont{Xu}},
  \bibinfo{journal}{Phys. Rev.} \textbf{\bibinfo{volume}{D100}},
  \bibinfo{pages}{034008} (\bibinfo{year}{2019}), \eprint{1903.00160}.

\bibitem[{\citenamefont{Yin et~al.}(2021)\citenamefont{Yin, Cui, Roberts, and
  Segovia}}]{Yin:2021uom}
\bibinfo{author}{\bibfnamefont{P.-L.} \bibnamefont{Yin}},
  \bibinfo{author}{\bibfnamefont{Z.-F.} \bibnamefont{Cui}},
  \bibinfo{author}{\bibfnamefont{C.~D.} \bibnamefont{Roberts}},
  \bibnamefont{and} \bibinfo{author}{\bibfnamefont{J.}~\bibnamefont{Segovia}},
  \bibinfo{journal}{Eur. Phys. J. C} \textbf{\bibinfo{volume}{81}},
  \bibinfo{pages}{327} (\bibinfo{year}{2021}), \eprint{2102.12568}.

\bibitem[{\citenamefont{Andronic et~al.}(2018)\citenamefont{Andronic,
  Braun-Munzinger, Redlich, and Stachel}}]{Andronic:2017pug}
\bibinfo{author}{\bibfnamefont{A.}~\bibnamefont{Andronic}},
  \bibinfo{author}{\bibfnamefont{P.}~\bibnamefont{Braun-Munzinger}},
  \bibinfo{author}{\bibfnamefont{K.}~\bibnamefont{Redlich}}, \bibnamefont{and}
  \bibinfo{author}{\bibfnamefont{J.}~\bibnamefont{Stachel}},
  \bibinfo{journal}{Nature} \textbf{\bibinfo{volume}{561}},
  \bibinfo{pages}{321} (\bibinfo{year}{2018}), \eprint{1710.09425}.

\bibitem[{\citenamefont{Sheng et~al.}(2024)\citenamefont{Sheng, Yang, Zou, and
  Hou}}]{Sheng:2022ssp}
\bibinfo{author}{\bibfnamefont{X.-L.} \bibnamefont{Sheng}},
  \bibinfo{author}{\bibfnamefont{S.-Y.} \bibnamefont{Yang}},
  \bibinfo{author}{\bibfnamefont{Y.-L.} \bibnamefont{Zou}}, \bibnamefont{and}
  \bibinfo{author}{\bibfnamefont{D.}~\bibnamefont{Hou}}, \bibinfo{journal}{Eur.
  Phys. J. C} \textbf{\bibinfo{volume}{84}}, \bibinfo{pages}{299}
  (\bibinfo{year}{2024}), \eprint{2209.01872}.

\bibitem[{\citenamefont{Wilson et~al.}(2012)\citenamefont{Wilson, Cloet, Chang,
  and Roberts}}]{Wilson:2011aa}
\bibinfo{author}{\bibfnamefont{D.~J.} \bibnamefont{Wilson}},
  \bibinfo{author}{\bibfnamefont{I.~C.} \bibnamefont{Cloet}},
  \bibinfo{author}{\bibfnamefont{L.}~\bibnamefont{Chang}}, \bibnamefont{and}
  \bibinfo{author}{\bibfnamefont{C.~D.} \bibnamefont{Roberts}},
  \bibinfo{journal}{Phys. Rev.} \textbf{\bibinfo{volume}{C85}},
  \bibinfo{pages}{025205} (\bibinfo{year}{2012}), \eprint{1112.2212}.

\bibitem[{\citenamefont{Eichmann
  et~al.}(2010{\natexlab{a}})\citenamefont{Eichmann, Alkofer, Krassnigg, and
  Nicmorus}}]{Eichmann:2009en}
\bibinfo{author}{\bibfnamefont{G.}~\bibnamefont{Eichmann}},
  \bibinfo{author}{\bibfnamefont{R.}~\bibnamefont{Alkofer}},
  \bibinfo{author}{\bibfnamefont{A.}~\bibnamefont{Krassnigg}},
  \bibnamefont{and} \bibinfo{author}{\bibfnamefont{D.}~\bibnamefont{Nicmorus}},
  \bibinfo{journal}{EPJ Web Conf.} \textbf{\bibinfo{volume}{3}},
  \bibinfo{pages}{03028} (\bibinfo{year}{2010}{\natexlab{a}}),
  \eprint{0912.2876}.

\bibitem[{\citenamefont{Eichmann
  et~al.}(2010{\natexlab{b}})\citenamefont{Eichmann, Alkofer, Krassnigg, and
  Nicmorus}}]{Eichmann:2009qa}
\bibinfo{author}{\bibfnamefont{G.}~\bibnamefont{Eichmann}},
  \bibinfo{author}{\bibfnamefont{R.}~\bibnamefont{Alkofer}},
  \bibinfo{author}{\bibfnamefont{A.}~\bibnamefont{Krassnigg}},
  \bibnamefont{and} \bibinfo{author}{\bibfnamefont{D.}~\bibnamefont{Nicmorus}},
  \bibinfo{journal}{Phys. Rev. Lett.} \textbf{\bibinfo{volume}{104}},
  \bibinfo{pages}{201601} (\bibinfo{year}{2010}{\natexlab{b}}),
  \eprint{0912.2246}.

\bibitem[{\citenamefont{Roberts
  et~al.}(2011{\natexlab{a}})\citenamefont{Roberts, Chang, Cloet, and
  Roberts}}]{Roberts:2011cf}
\bibinfo{author}{\bibfnamefont{H.~L.~L.} \bibnamefont{Roberts}},
  \bibinfo{author}{\bibfnamefont{L.}~\bibnamefont{Chang}},
  \bibinfo{author}{\bibfnamefont{I.~C.} \bibnamefont{Cloet}}, \bibnamefont{and}
  \bibinfo{author}{\bibfnamefont{C.~D.} \bibnamefont{Roberts}},
  \bibinfo{journal}{Few Body Syst.} \textbf{\bibinfo{volume}{51}},
  \bibinfo{pages}{1} (\bibinfo{year}{2011}{\natexlab{a}}), \eprint{1101.4244}.

\bibitem[{\citenamefont{Chen et~al.}(2012)\citenamefont{Chen, Chang, Roberts,
  Wan, and Wilson}}]{Chen:2012qr}
\bibinfo{author}{\bibfnamefont{C.}~\bibnamefont{Chen}},
  \bibinfo{author}{\bibfnamefont{L.}~\bibnamefont{Chang}},
  \bibinfo{author}{\bibfnamefont{C.~D.} \bibnamefont{Roberts}},
  \bibinfo{author}{\bibfnamefont{S.}~\bibnamefont{Wan}}, \bibnamefont{and}
  \bibinfo{author}{\bibfnamefont{D.~J.} \bibnamefont{Wilson}},
  \bibinfo{journal}{Few Body Syst.} \textbf{\bibinfo{volume}{53}},
  \bibinfo{pages}{293} (\bibinfo{year}{2012}), \eprint{1204.2553}.

\bibitem[{\citenamefont{Roberts
  et~al.}(2011{\natexlab{b}})\citenamefont{Roberts, Bashir, Gutierrez-Guerrero,
  Roberts, and Wilson}}]{Roberts:2011wy}
\bibinfo{author}{\bibfnamefont{H.~L.~L.} \bibnamefont{Roberts}},
  \bibinfo{author}{\bibfnamefont{A.}~\bibnamefont{Bashir}},
  \bibinfo{author}{\bibfnamefont{L.~X.} \bibnamefont{Gutierrez-Guerrero}},
  \bibinfo{author}{\bibfnamefont{C.~D.} \bibnamefont{Roberts}},
  \bibnamefont{and} \bibinfo{author}{\bibfnamefont{D.~J.}
  \bibnamefont{Wilson}}, \bibinfo{journal}{Phys. Rev.}
  \textbf{\bibinfo{volume}{C83}}, \bibinfo{pages}{065206}
  (\bibinfo{year}{2011}{\natexlab{b}}), \eprint{1102.4376}.

\bibitem[{\citenamefont{Takahashi et~al.}(2026)}]{ParticleDataGroup:2026aaa}
\bibinfo{author}{\bibfnamefont{F.}~\bibnamefont{Takahashi}}
  \bibnamefont{et~al.} (\bibinfo{collaboration}{Particle Data Group}),
  \bibinfo{journal}{Int. J. Mod. Phys. A} \textbf{\bibinfo{volume}{41}},
  \bibinfo{pages}{2630011} (\bibinfo{year}{2026}).

\bibitem[{\citenamefont{Qin et~al.}(2019)\citenamefont{Qin, Roberts, and
  Schmidt}}]{Qin:2019hgk}
\bibinfo{author}{\bibfnamefont{S.-x.} \bibnamefont{Qin}},
  \bibinfo{author}{\bibfnamefont{C.~D.} \bibnamefont{Roberts}},
  \bibnamefont{and} \bibinfo{author}{\bibfnamefont{S.~M.}
  \bibnamefont{Schmidt}}, \bibinfo{journal}{Few Body Syst.}
  \textbf{\bibinfo{volume}{60}}, \bibinfo{pages}{26} (\bibinfo{year}{2019}),
  \eprint{1902.00026}.

\bibitem[{\citenamefont{Bahtiyar et~al.}(2020)\citenamefont{Bahtiyar, Can,
  Erkol, Gubler, Oka, and Takahashi}}]{Bahtiyar:2020uuj}
\bibinfo{author}{\bibfnamefont{H.}~\bibnamefont{Bahtiyar}},
  \bibinfo{author}{\bibfnamefont{K.~U.} \bibnamefont{Can}},
  \bibinfo{author}{\bibfnamefont{G.}~\bibnamefont{Erkol}},
  \bibinfo{author}{\bibfnamefont{P.}~\bibnamefont{Gubler}},
  \bibinfo{author}{\bibfnamefont{M.}~\bibnamefont{Oka}}, \bibnamefont{and}
  \bibinfo{author}{\bibfnamefont{T.~T.} \bibnamefont{Takahashi}},
  \bibinfo{journal}{Phys. Rev. D} \textbf{\bibinfo{volume}{102}},
  \bibinfo{pages}{054513} (\bibinfo{year}{2020}), \eprint{2004.08999}.

\bibitem[{\citenamefont{Li et~al.}(2025{\natexlab{b}})\citenamefont{Li, Gui,
  Qin, Sun, and Liang}}]{Li:2022vbc}
\bibinfo{author}{\bibfnamefont{J.-B.} \bibnamefont{Li}},
  \bibinfo{author}{\bibfnamefont{L.-C.} \bibnamefont{Gui}},
  \bibinfo{author}{\bibfnamefont{W.}~\bibnamefont{Qin}},
  \bibinfo{author}{\bibfnamefont{W.}~\bibnamefont{Sun}}, \bibnamefont{and}
  \bibinfo{author}{\bibfnamefont{J.}~\bibnamefont{Liang}},
  \bibinfo{journal}{Chin. Phys. C} \textbf{\bibinfo{volume}{49}},
  \bibinfo{pages}{063103} (\bibinfo{year}{2025}{\natexlab{b}}),
  \eprint{2211.04713}.

\bibitem[{\citenamefont{Suenaga and Oka}(2025)}]{Suenaga:2024vwr}
\bibinfo{author}{\bibfnamefont{D.}~\bibnamefont{Suenaga}} \bibnamefont{and}
  \bibinfo{author}{\bibfnamefont{M.}~\bibnamefont{Oka}},
  \bibinfo{journal}{Phys. Rev. D} \textbf{\bibinfo{volume}{111}},
  \bibinfo{pages}{074032} (\bibinfo{year}{2025}), \eprint{2411.12172}.

\end{thebibliography}

\end{document}